\pdfoutput=1
\documentclass[cernpreprint,atlasdraft=false,texlive=2016,UKenglish,texmf]{atlasdoc}
 
\usepackage[subcaption]{atlaspackage}
\usepackage{atlasbiblatex}
 
\usepackage{atlasphysics}
 
\graphicspath{{logos/}{figures/}}
 
\usepackage{ANA-HIGG-2018-43-PAPER-defs}
 
\usepackage{todonotes}

\AtlasTitle{Evidence for Higgs boson decays to a low-mass dilepton system and a photon in $pp$ collisions at $\sqrt{s} = 13$~TeV with the ATLAS detector}
 
\AtlasAbstract{
A search for the Higgs boson decaying into a photon and a pair of electrons or muons with an invariant mass $m_{\ell\ell} < 30$~GeV is presented.
The analysis is performed using 139~fb$^{-1}$ of proton--proton collision data, produced by the LHC at a centre-of-mass energy of 13~TeV and collected by the ATLAS experiment. Evidence for the $H \to \ell\ell\gamma$ process is found with a significance of 3.2 over the background-only hypothesis, compared to an expected significance of 2.1 for the Standard Model prediction.
The best-fit value of the signal-strength parameter, defined as the ratio of the observed signal yield to the one expected in the Standard Model, is $\mu = 1.5 \pm 0.5$. The Higgs boson production cross-section times the $H \rightarrow\ell\ell\gamma$ branching ratio for $m_{\ell\ell} < 30$~GeV is determined to be $8.7~^{+2.8}_{-2.7}$~fb.
}
 
\author{The ATLAS Collaboration}
 
\AtlasRefCode{HIGG-2018-43}
 
\PreprintIdNumber{CERN-EP-2021-035}

\AtlasJournalRef{Phys. Lett. B 819 (2021) 136412}
\AtlasDOI{10.1016/j.physletb.2021.136412}

\hypersetup{pdftitle={ATLAS document},pdfauthor={The ATLAS Collaboration}}
 
\begin{document}
 
\maketitle

\section{Introduction}
\label{sec:intro}
In July 2012, the ATLAS and CMS Collaborations at the CERN Large Hadron
Collider (LHC) announced the discovery
of a new particle with a mass of approximately
$125~\GeV$~\cite{HIGG-2012-27, CMS-HIG-12-028}.
The observed properties of the particle, such as its couplings to Standard Model (SM) elementary
particles, its spin and its parity, are so far consistent with the predictions
for the SM Higgs boson~\cite{HIGG-2013-17, CMS-HIG-14-018, HIGG-2015-07, CMS-HIG-17-031,HIGG-2018-57}.
 
Measurements of rare decays of the Higgs boson, such as \hllg where $\ell$ is an electron or muon, can probe coupling modifications introduced by possible extensions to the SM~\cite{Kachanovich:2020xyg}. In addition, such three-body Higgs boson decays can be used to probe $CP$-violation in the Higgs sector~\cite{Korchin:2014kha,Chen:2014ona}.
 
\begin{figure}[h!]
\centering
\begin{tabular}{cccc}
\includegraphics[width=0.22\textwidth]{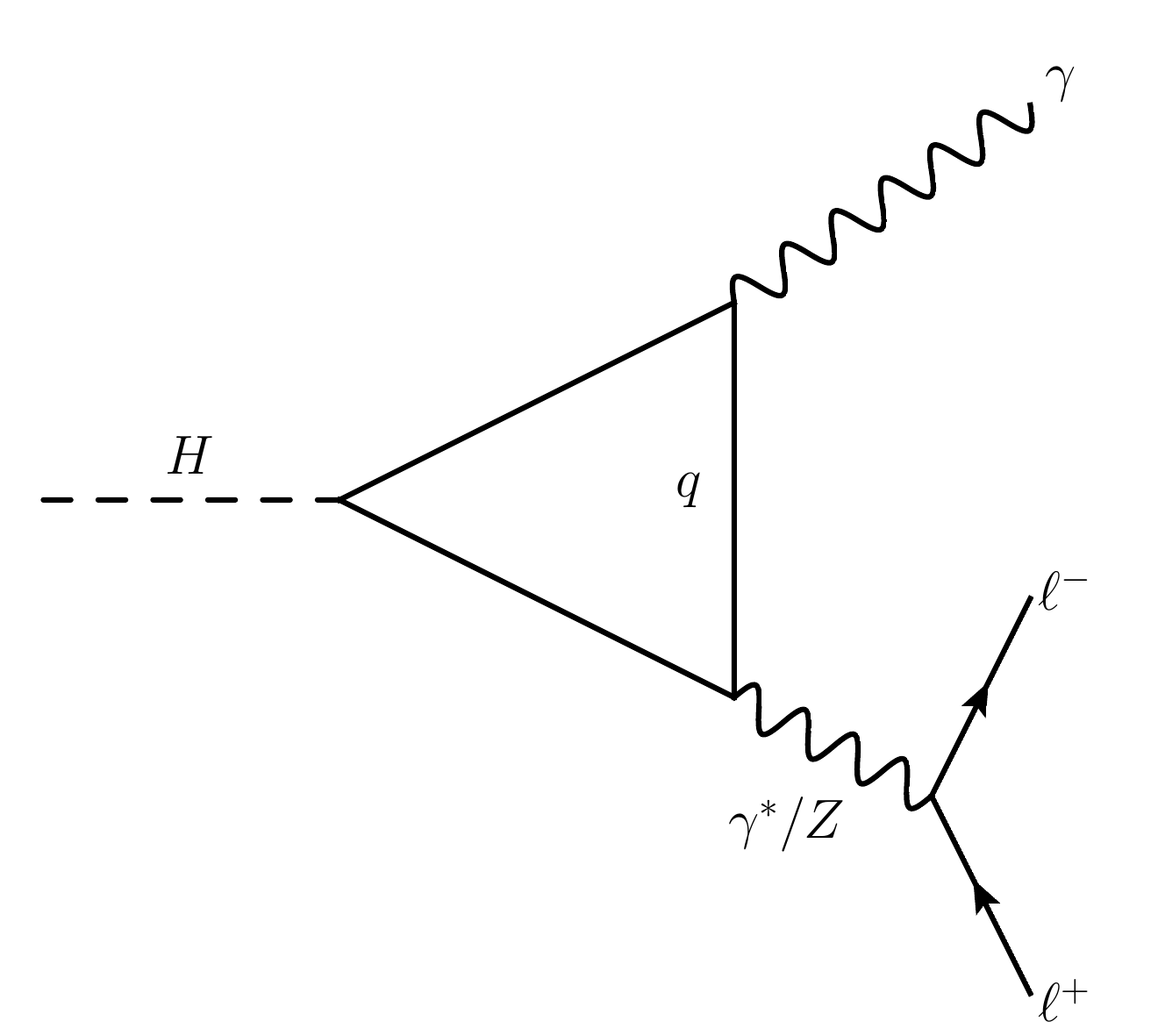} &
\includegraphics[width=0.22\textwidth]{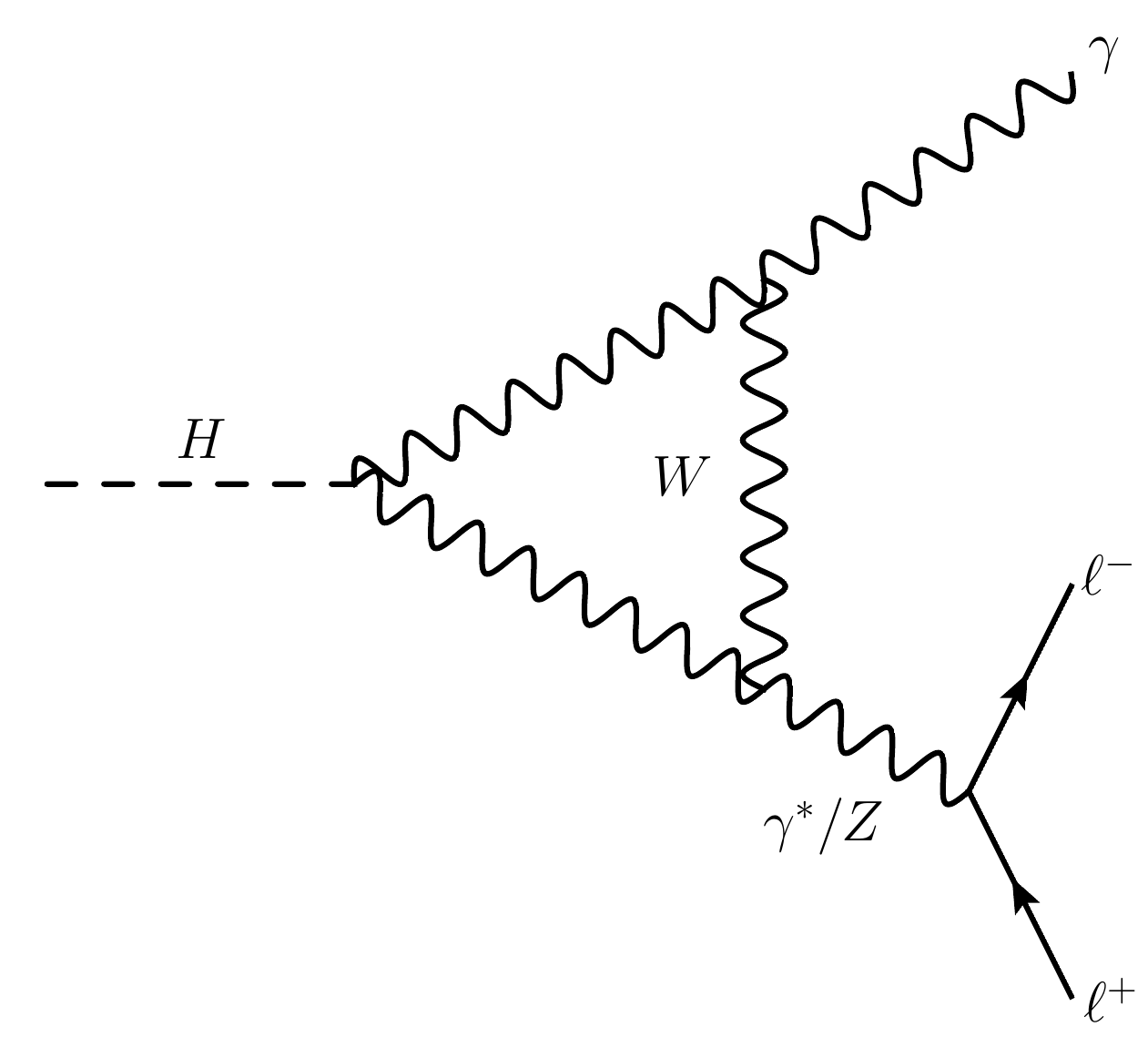} &
\includegraphics[width=0.22\textwidth, trim = {0 -1cm 0 0cm}]{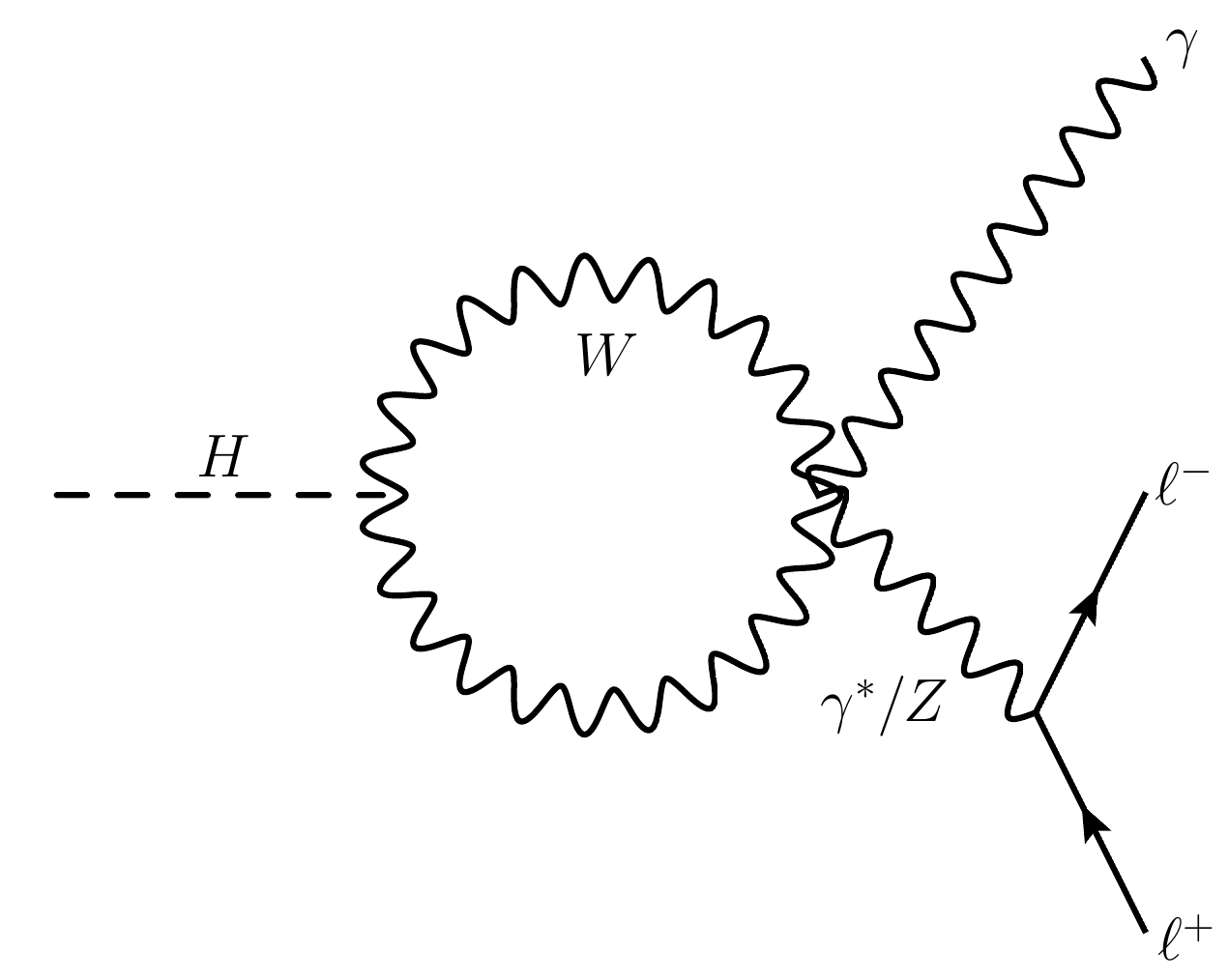}  &
\includegraphics[width=0.2\textwidth, trim = {0 -1cm 0 0cm}]{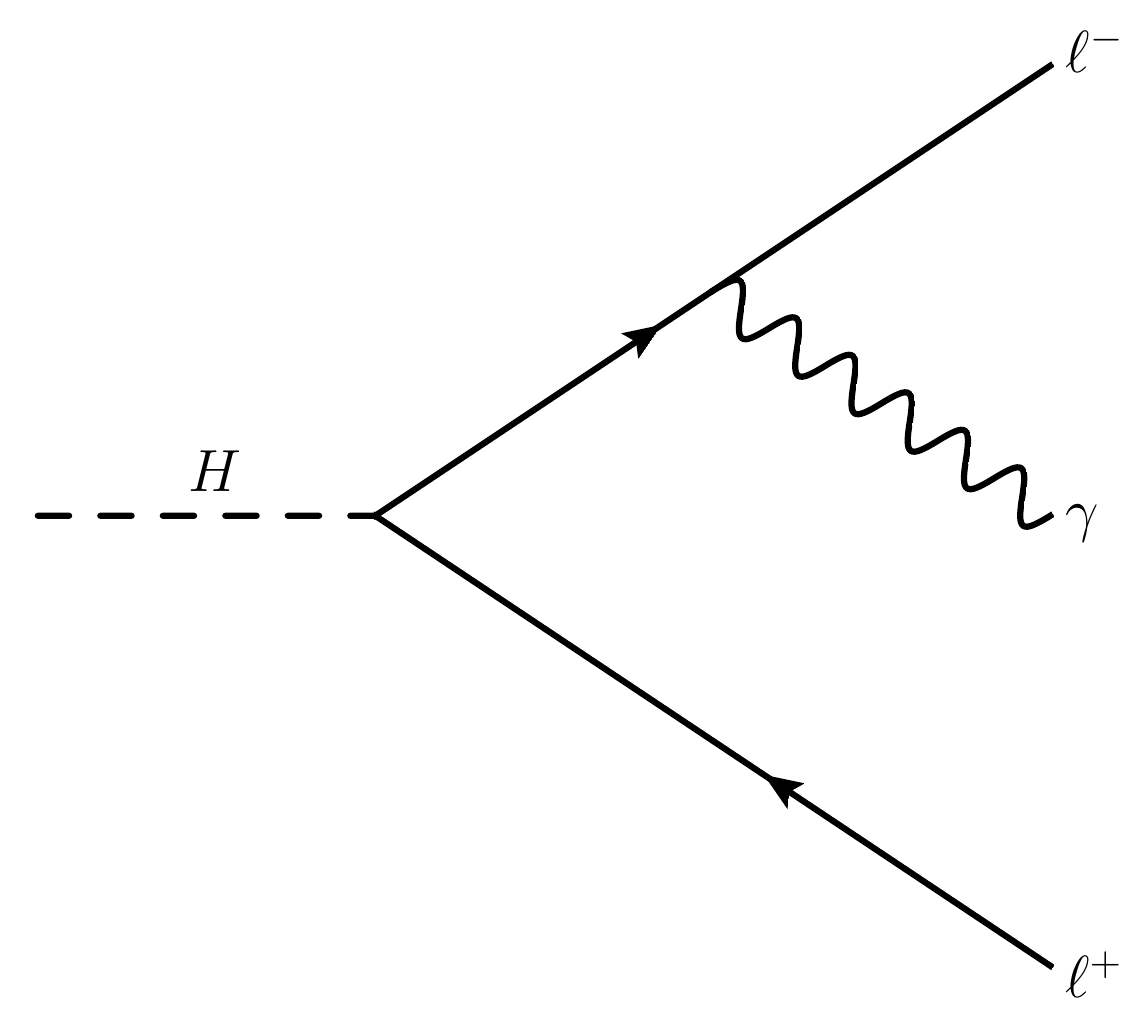} \\
(a) & (b) & (c) & (d)\\
\end{tabular}
\caption{Representative Feynman diagrams of the \hllg process.
}
\label{fig:feynmans}
\end{figure}

Multiple processes contribute to the \hllg\ decay: Dalitz decays involving a $Z$ boson or a virtual photon ($\gamma^*$) (Figure~\ref{fig:feynmans}(a--c)), as well as the decay of the Higgs boson to two leptons and a photon from final-state radiation (FSR) (Figure~\ref{fig:feynmans}(d)). Their respective fractions depend on the invariant mass of the dilepton pair, \mll. In this analysis only low-mass dilepton pairs with $\mll < 30$~\GeV\ are considered. This region is completely dominated by the decay through $\gamma^*$~\cite{Chen:2012ju,Passarino:2013nka,Kachanovich:2020xyg}. The contributions of the other processes and interferences are negligible.

Based on a data sample of proton--proton ($pp$) collisions at \sqtt with an integrated luminosity of 35.9~\ifb, the CMS Collaboration reported a 95\%~CL upper limit on the production cross-section times branching ratio  for the low-$m_{\mu\mu}$ \hmumuy process of 4.0 times the SM prediction~\cite{CMSZylow-2018}.
In addition, both the ATLAS and CMS Collaborations carried out searches at \sqtt for the closely related \htoZg process~\cite{CMSZylow-2018,HIGG-2018-42}.
The CMS Collaboration also searched for the low-\mll \hllg process in the dimuon and dielectron channels in $pp$ collisions at $\sqrt{s}=8~\TeV$~\cite{Khachatryan:2015lga}.

This paper describes a search for \heeg and \hmumug decays with \mll $<$ 30~\GeV.
When the invariant mass of the two electrons is low and the transverse momentum of the dielectron system is high, their electromagnetic showers can overlap in the calorimeter.
Therefore, the search for \eeg final states requires the development of dedicated electron trigger and identification algorithms.

The search uses $pp$ collision data at  \sqtt recorded
with the ATLAS detector during Run~2 of the LHC between 2015 and 2018, corresponding to a total
integrated luminosity of \lumithirteentev~\invfb.
The sensitivity of the search is enhanced by dividing the selected events into mutually exclusive categories, according to the event topology and lepton flavour.
The dominant background is the irreducible non-resonant production of \llg.
After event categorisation, the signal yield is extracted by a simultaneous fit of parametric functions to the reconstructed \llg invariant mass (\mllg) distributions in all categories.


\section{ATLAS detector}
\label{sec:detector}
The ATLAS detector \cite{PERF-2007-01} covers nearly the
entire solid angle around the collision point.\footnote{The ATLAS experiment uses a
right-handed coordinate system with its origin at the nominal interaction point (IP) in the centre
of the detector and the $z$-axis along the beam pipe. The $x$-axis points from the IP to the centre of the
LHC ring, and the $y$-axis points upward. Cylindrical coordinates $(r,\phi)$ are used in the transverse plane.
The polar angle ($\theta$) is measured from the positive $z$-axis and the azimuthal angle ($\phi$) is measured from the positive $x$-axis in the transverse plane.
The pseudorapidity is defined as $\eta=-\ln\tan(\theta/2)$. Angular distance is measured in units of $\Delta R = \sqrt{(\Delta\eta)^2 + (\Delta\phi)^2}$.} It consists of an inner tracking detector (ID) surrounded by a thin superconducting solenoid
providing a \SI{2}{\tesla} axial magnetic field, electromagnetic (EM) and hadron calorimeters, and a muon spectrometer (MS).
The ID covers the pseudorapidity range \(|\eta| < 2.5\).
It consists of silicon pixel, silicon microstrip, and transition radiation tracking detectors.
The silicon pixel detector provides up to four measurements per track. The insertable B-layer (IBL)~\cite{ATLASDET-IBL,PIX-2018-001} constitutes the innermost layer at a radius of 33.3 mm. It surrounds the beam pipe, which has an inner radius of 23.5~mm.
Lead/liquid-argon (LAr) sampling calorimeters provide EM energy measurements
with high granularity.
A steel/scintillator-tile hadron calorimeter covers the central pseudorapidity range (\(|\eta| < 1.7\)).
The endcap and forward regions are instrumented with LAr calorimeters
for both the EM and hadronic energy measurements up to \(|\eta| = 4.9\).
For \(|\eta| < 2.5\), the EM calorimeter is divided into three longitudinal layers, which are finely segmented in $\eta$ and $\phi$, particularly in the first layer. This segmentation allows the measurement of the lateral and longitudinal shower profile, and the calculation of shower shapes~\cite{EGAM-2018-01} used for particle identification and background rejection. The longitudinal segmentation of the EM calorimeter is also exploited to calibrate the energy response of electron and photon candidates~\cite{EGAM-2018-01}.
The MS comprises separate trigger and
high-precision tracking chambers measuring the deflection of muons in
a magnetic field generated by superconducting air-core toroids. The
precision chamber system covers the region $|\eta| < 2.7$ with three
layers of monitored drift tubes, complemented by cathode-strip
chambers in the forward region, where the background is highest. The
muon trigger system covers the range $|\eta| < 2.4$ with
resistive-plate chambers in the barrel, and thin-gap chambers in the
endcap regions.

A two-level trigger system~\cite{TRIG-2016-01} was used during the \sqtt~data-taking period.
The first-level trigger (L1) is implemented
in hardware and uses a subset of the detector information. This is followed by a software-based
high-level trigger which runs algorithms similar to those in the offline reconstruction software, reducing the event rate to
approximately 1~kHz from the maximum L1 rate of 100~kHz.
 

\section{Data and simulated event samples}
\label{sec:datamcsamples}
The analysed $pp$ collision data
correspond to the full recorded LHC Run-2 data set at $\sqrt{s} = 13$~\TeV\, with an integrated luminosity of
\lumithirteentev~\invfb after data quality requirements~\cite{DAPR-2018-01}.
The events were collected with a combination of single-lepton, dilepton, diphoton, and lepton+photon triggers.
Standard electron triggers, which require narrow, isolated EM energy clusters, are efficient if the two electrons in the event have a very small angular separation $\Delta R_{ee}$, as the two electrons produce a cluster that is similar to that of a single electron. For $\Delta R_{ee} > 0.1$, two separate EM clusters are typically produced.
For the 2017 and 2018 data-taking periods, a dedicated trigger was introduced
which requires at least one photon with $\pt>35~\GeV$ and at least one EM cluster with  $\pt>25~\GeV$ matched to at least one ID track. The EM cluster must have low hadronic leakage, but no requirement is imposed on the width of the shower, which allows this new trigger to recover a significant fraction of the signal with $0.025 < \Delta R_{ee} < 0.1$.
More details are provided in Ref.~\cite{TRIG-2018-05}.
The overall trigger efficiencies for the \heeg and \hmumug processes are $98\%$ and $96\%$, respectively ($97\%$ combined), relative to the event selection
discussed in Section~\ref{sec:evtselection}.
The uncertainty in the trigger efficiency is negligible.
 
Samples of simulated Monte Carlo (MC) events are crucial for the optimisation of the search strategy and the modelling of the signal and background processes.
Simulated \htoystary signal and \hgg background events are used to parameterise these processes, and simulated non-Higgs \llg events are used to choose analytic functional forms to describe the non-resonant background.
The generated MC events, unless stated otherwise, were processed with the full ATLAS detector simulation~\cite{ATLAS-SIM}
based on $\GEANT4$~\cite{Geant4}.
The effect of multiple $pp$ interactions in the same and neighbouring bunch
crossings (pile-up) was included by overlaying minimum-bias events simulated
with \PYTHIAV{8.186}~\cite{Sjostrand:2007gs} using the NNPDF2.3LO parton distribution function (PDF) set~\cite{Ball:2012cx}  and the A3 set~\cite{ATL-PHYS-PUB-2016-017} of tuned parameters.
The MC events were weighted to reproduce the distribution of the number of interactions per bunch crossing observed in the data.
 
Higgs boson production in the gluon--gluon fusion (\ggF) and vector-boson fusion (\VBF) production modes, as well as in quark-initiated associated production with a $W$ or $Z$ boson (\qqVH) or with two top quarks (\ttH) was modelled with the \POWHEGBOXV{v2}  MC event generator~\cite{Nason:2009ai,Alioli:2010xd,Nason:2004rx,Frixione:2007vw,Hartanto:2015uka}. \POWHEGBOXV{v2} was interfaced with \PYTHIAV{8}~\cite{Sjostrand:2007gs} to simulate the \htoystary and $H\to\gamma\gamma$ decays. \PYTHIA also provides parton showering, hadronisation and the underlying event. The PDF4LHC15 PDF set~\cite{Butterworth:2015oua} was used, except for \ttH, where the NNPDF3.0nlo PDF set~\cite{Ball:2012cx} was used.
For the \ggF process, the \POWHEG NNLOPS program~\cite{Hamilton:2013fea,Hamilton:2015nsa} achieves next-to-next-to-leading-order (NNLO) accuracy in QCD for inclusive observables after reweighting the Higgs boson rapidity spectrum~\cite{Catani:2007vq}.
The simulation reaches next-to-leading-order (NLO) accuracy in QCD for the \VBF, \VH, and \ttH processes. For \VH, the  MiNLO technique~\cite{Cullen:2011ac,Hamilton:2012rf,Luisoni:2013kna} was applied.
No simulated samples are available for gluon-initiated associated production with a $Z$ boson (\ggZH) and associated production with two $b$-quarks (\bbH). Their minor contribution, 1\% of the expected Higgs boson production, is modelled using the acceptances from the \qqZH and \ggF samples, respectively.

Alternative \hyy samples are considered for the \ggF and \VBF processes, where \HERWIGV{7}~\cite{Bellm:2015jjp} was used instead of \PYTHIAV{8} to provide parton showering. Weights were calculated by comparing the generator-level Higgs boson transverse momentum and jet distributions in these samples with the distributions obtained from the nominal \hyy samples. The  weights are applied to the \hysy \ggF and \VBF samples to estimate the effect of varying the parton shower and underlying event.

The mass of the Higgs boson was set in the
simulation to $m_H = 125~\GeV$ and its width to $\Gamma_H =4.1~\MeV$~\cite{deFlorian:2016spz}.
The Higgs mass peak position in the simulation was corrected to account for the small difference relative to the measured value of 125.09~\GeV~\cite{HIGG-2014-14} (see Section~\ref{sec:sigbkgmodel}); otherwise the effect on the kinematics is neglected.

The signal and \hyy samples were normalised with the best available theoretical calculations of the corresponding SM production cross-sections at $\mh = 125.09$~\GeV, which are available at next-to-next-to-next-to-leading order (N3LO) in QCD with NLO electroweak (EW) corrections for \ggF~\cite{deFlorian:2016spz,Anastasiou:2015ema,Anastasiou:2016cez,Dulat:2018rbf,Harlander:2009mq,Harlander:2009bw,Harlander:2009my,Pak:2009dg,Actis:2008ug,Actis:2008ts,Anastasiou:2008tj,Aglietti:2004nj,Bonetti:2018ukf},
at approximate-NNLO in QCD with NLO EW corrections for \VBF~\cite{deFlorian:2016spz,Ciccolini:2007jr,Ciccolini:2007ec,Bolzoni:2010xr}, at NNLO in QCD with NLO EW corrections for \qqVH~\cite{deFlorian:2016spz,Ciccolini:2003jy,Brein:2003wg,Brein:2011vx,Altenkamp:2012sx,Denner:2014cla,Brein:2012ne,Harlander:2014wda, Harlander:2018yio}
and at NLO in QCD with NLO electroweak corrections for \ttH~\cite{Beenakker:2002nc,Dawson:2003zu, Yu:2014cka,Frixione:2015zaa,deFlorian:2016spz}.
The accuracy of the calculation is NNLO in QCD for \bbH~\cite{Dawson:2003kb,Dittmaier:2003ej,Harlander:2011aa}, and NLO in QCD with next-to-leading-logarithm (NLL) corrections for \ggZH~\cite{deFlorian:2016spz,Altenkamp:2012sx}.

There are multiple calculations of the \hllg branching ratios for different slices of phase space~\cite{Abbasabadi:1996ze,Firan:2007tp,Chen:2012ju,Dicus:2013ycd,Passarino:2013nka,Dicus:2014mra,Kachanovich:2020xyg}.
The \hllg branching ratios used in this analysis are $\mathcal{B}(H \rightarrow ee\gamma) = 7.20 \times 10^{-5}$ and $\mathcal{B}(H \rightarrow \mu\mu\gamma) = 3.42 \times 10^{-5}$, as estimated with \PYTHIAV{8} for \mll $<$ 30 \GeV. Their sum corresponds to $\sim$5\% of the \hyy\ branching ratio.
When extrapolated to a common phase space, the \PYTHIAV{8} estimate agrees with the predictions of Refs.~\cite{Firan:2007tp,Dicus:2014mra}
within a relative difference of 3\%.
Currently, no theoretical uncertainty calculation exists for the low-\mll \hllg branching ratio.
Therefore, the theoretical uncertainties in the \hyy\ and \htoZg branching ratios are considered~\cite{Passarino:2013nka}, and the larger of the two uncertainties is used, which corresponds to the 5.8\% relative uncertainty in the \htoZg branching ratio~\cite{deFlorian:2016spz}.

The background originates mainly from non-resonant \llg production.
Events were simulated with the
\SHERPAV{2.2.8}~\cite{Bothmann:2019yzt} generator based on LO matrix elements for \llg
production with up to three additional partons and using the NNPDF3.0
PDF set~\cite{Ball:2014uwa}. The \SHERPA{} simulation includes parton shower, fragmentation and underlying-event modelling.
As the statistical uncertainties in the simulated \llg background samples are a limiting factor when studying the background modelling
at the level required by the small signal-to-background ratio, a procedure was developed to generate significantly larger samples.
These  are based on events generated using the \SHERPAV{2.2.8} configuration described above with object efficiencies
approximated by parameterisations rather than using the full ATLAS detector simulation and reconstruction software.
The parameterisations, extracted from fully simulated MC samples,
reproduce the reconstruction and selection efficiencies of detector-level objects via event weighting.
Comparisons with a sample that underwent the full ATLAS detector simulation, and whose effective luminosity is greater than the data luminosity, show good agreement within the statistical uncertainties.

 
\section{Object selection}
\label{sec:objselection}
Events are required to have a collision vertex associated with at least two tracks with transverse momentum $\pt > 0.5~\GeV$ each. In the case of multiple vertices, the vertex with the largest $\sum \pt^2$ of the associated tracks is considered to be the primary vertex.
 
Muon candidates are obtained by matching high-quality tracks in the MS and ID.
Standalone MS tracks are used to extend the muon reconstruction beyond the ID acceptance to $2.5 < |\eta| < 2.7$.
Muon candidates are required to satisfy the \textit{medium} identification criteria~\cite{Aad:2016jkr, Aad:2020gmm}, be within $|\eta|<2.7$ and have $\pt > 3$~\GeV.
Muon candidates with an associated ID track must be matched to the primary vertex by having a longitudinal impact parameter $\Delta\zzero$ that satisfies $\zzsint < 0.5$~mm, where $\theta$ is the polar angle of the track. The significance of the transverse impact parameter $\dzero$ calculated relative to the measured beam-line position is required to be $|\dzero| / \sigma_{\dzero} < 3$, where $\sigma_{\dzero}$ is the uncertainty in $\dzero$.
A subset of the muon candidates are also required to be isolated from additional activity in the tracking detector and in the calorimeters, using a loose isolation selection~\cite{Aad:2016jkr, Aad:2020gmm}.
The efficiency of the muon reconstruction and identification, as well as the momentum calibration, including the associated systematic uncertainties, is estimated as described in Refs.~\cite{Aad:2016jkr, Aad:2020gmm}.

Photon and electron candidates are reconstructed from energy clusters in the calorimeters which are formed
using an algorithm based on dynamical, topological cell-clustering~\cite{EGAM-2018-01}.
Candidates in the transition region between the barrel and endcap EM calorimeters, $1.37 < |\eta| < 1.52$, are excluded.
The performance of the electron and photon reconstruction, including the associated systematic uncertainties, is studied in Ref.~\cite{EGAM-2018-01}.

Photon candidates can be either unconverted or converted. In the latter case, a track or conversion vertex is matched to the EM cluster. Photon candidates are required to satisfy $\pt > 20$~\GeV\ and $|\eta|<2.37$, and pass the \textit{tight} identification requirements~\cite{PERF-2017-02,EGAM-2018-01} as well as a loose isolation selection~\cite{EGAM-2018-01}.

The energy of the EM clusters associated with the photon and electron candidates is corrected in successive steps using a combination of simulation-based and data-driven correction factors~\cite{EGAM-2018-01}.
The simulation-based calibration regression is optimised separately for electrons, unconverted photons and converted photons.
The resolution of the energy response is corrected in the simulation to match the resolution observed in data using $Z\to ee$  events, by smearing the electron energy such that the width of the simulated $Z$ boson peak matches the width observed in data.

Because of the event kinematics of the signal process, it is common for the energy deposits of the two electrons in the EM calorimeter to be reconstructed as a single cluster.
Therefore, two types of electron candidates are defined, each with its own selection criteria: one in which a topological cluster of energy deposits is associated with one selected ID track (\textit{resolved} electron~\cite{PERF-2017-01}), representing a single electron, and one in which a topological cluster is associated with two selected ID tracks (\textit{merged ee}), representing a merged electron pair.
Each ID track considered must satisfy $\zzsint < 0.5$~mm and $|\dzero| / \sigma_{\dzero} < 5$.
Resolved electron candidates must satisfy the \textit{medium} likelihood-based identification criteria~\cite{PERF-2017-01, EGAM-2018-01}, have $\pt > 4.5$~\GeV\ and be within $|\eta|<2.47$.
A subset of the resolved electron candidates are also required to pass a loose isolation requirement~\cite{EGAM-2018-01}.

ID tracks considered for merged electrons must have opposite charge, $\pt>5$~\GeV, $|\eta|<2.5$, and at least seven hits in the pixel and microstrip detectors combined.
To suppress backgrounds from converted photons, tracks must also have a hit in the innermost pixel layer, and merged-electron candidates are rejected if they match a conversion vertex with a radius larger than 20~mm whose momentum agrees better with the cluster energy than the momentum of the track that geometrically matches the cluster best.
Because the kinematic behaviour of merged electrons in the calorimeter most closely resembles
a photon converting in the material close to the interaction vertex,
the energy of the merged-$ee$ object is calibrated as a converted photon with a conversion radius set to $r_{\textrm{conv}}=30$~mm calculated relative to the measured beam-line position.
As presented in Figure~\ref{fig:merged_perf}(a), this treatment is found to induce minimal bias in the dielectron energy measurement.
To cover remaining differences between the simulated detector response to converted photons and merged-$ee$ objects, the difference between the squares of the energy resolutions for converted photons and merged-$ee$ objects in the MC simulation is treated as the square of an additional energy resolution uncertainty.
The four-momentum of the merged-$ee$ candidate is constructed using the calibrated energy and the direction and invariant mass obtained from the vertex reconstructed from the two electron-track candidates~\cite{Fruhwirth:1987fm}.
 
Merged-$ee$ candidates are required to have  $|\eta|<2.37$ (excluding $1.37 < |\eta| < 1.52$) and $\pt>20$~\GeV, and satisfy dedicated
identification requirements, as the standard electron criteria have a low efficiency for objects with closely spaced energy deposits or broader EM showers. For the merged-$ee$ identification, a multivariate discriminator is trained to separate the $\gamma^*\to ee$ signal objects from jets or single electrons.
The input variables for the training include shower shape variables~\cite{PERF-2017-01}, the information provided by the transition radiation tracker~\cite{ATLAS-CONF-2011-128}, and the kinematic information from the cluster and ID tracks.
Merged-$ee$ candidates are also required to pass a tight isolation requirement~\cite{EGAM-2018-01}.
A combined efficiency of $\sim$50\%  for the merged-$ee$ identification and isolation is achieved for $H \to \gamma^*\gamma \to ee \gamma$  events.
Since photons with a relatively small conversion radius offer a signature similar to that of merged-$ee$ objects, the merged-$ee$ identification  and isolation efficiency is measured in data using a tag-and-probe method
with FSR photons from $Z$ boson radiative decays ($Z\to\ell\ell\gamma$).
Candidate $Z\to\ell\ell\gamma$ events are selected similarly to those in Ref.~\cite{PERF-2017-02}.
Only two-track converted photons with $r_{\textrm{conv}}<160$~mm, corresponding to conversions inside the silicon pixel detector volume, are considered.
The $Z\to\ell\ell\gamma$ and background yields are estimated from a fit to the $\mllg$ distribution in data.
The extracted efficiencies are compared with the efficiencies estimated from simulated $Z\to\ell\ell\gamma$ events as shown in Figure~\ref{fig:merged_perf}(b) for photons with $|\eta|<0.8$.
The resulting $\pt$- and $\eta$-dependent data/simulation scale factors are between 0.9 and 1.1 and are used to correct the identification and isolation efficiencies of the simulated $H \to \gamma^*\gamma \to ee \gamma$  events.
The statistical uncertainties of the scale factors are taken into account.
In addition, a systematic uncertainty is assessed for the background modelling by varying the selection criteria of the $\mllg$ background template.
The total uncertainty reaches 2\% for $20 < \pT < 30~\GeV$ and 9\% for $\pT>50~\GeV$.
Figure~\ref{fig:merged_perf}(b) also shows a comparison of the extracted efficiencies with efficiencies in simulated $H \to \gamma^*\gamma \to ee \gamma$ events where an additional generator-level requirement of $|\Delta\eta_{ee}|<0.003$ is used. This requirement is applied in order to better match the signal signature to the converted photon signature in the detector; approximately 70\% of the merged-$ee$ objects in the signal sample pass it.
The efficiencies of converted photons and the merged-$ee$ objects agree within 10\% for $\pT<30~\GeV$ and within 5\% for $\pT>30~\GeV$ in the entire $\eta$ range.

\begin{figure}[htb]
\begin{center}
\begin{tabular}{cc}
\includegraphics[width=0.465\textwidth]{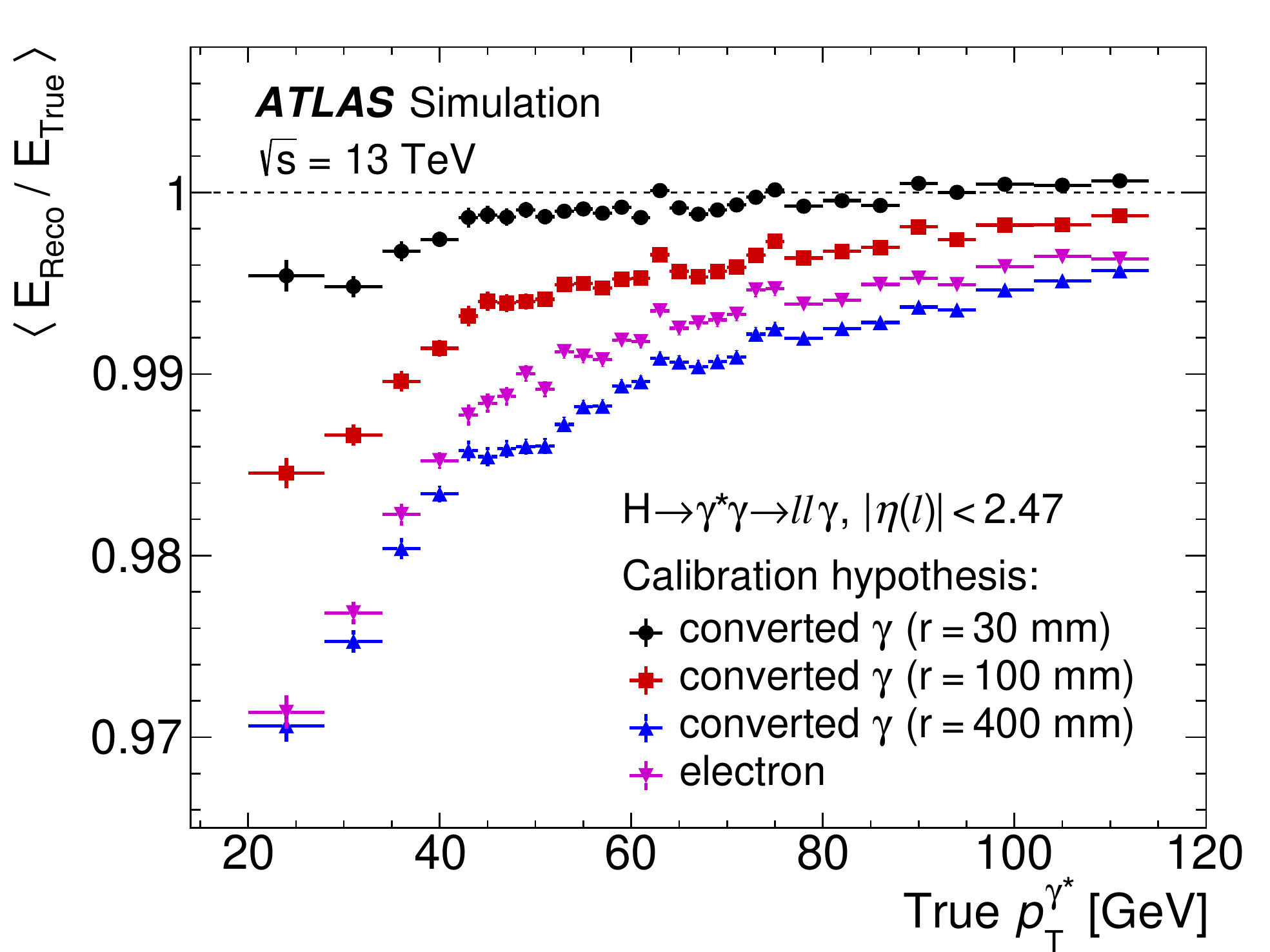} &
\includegraphics[width=0.465\textwidth]{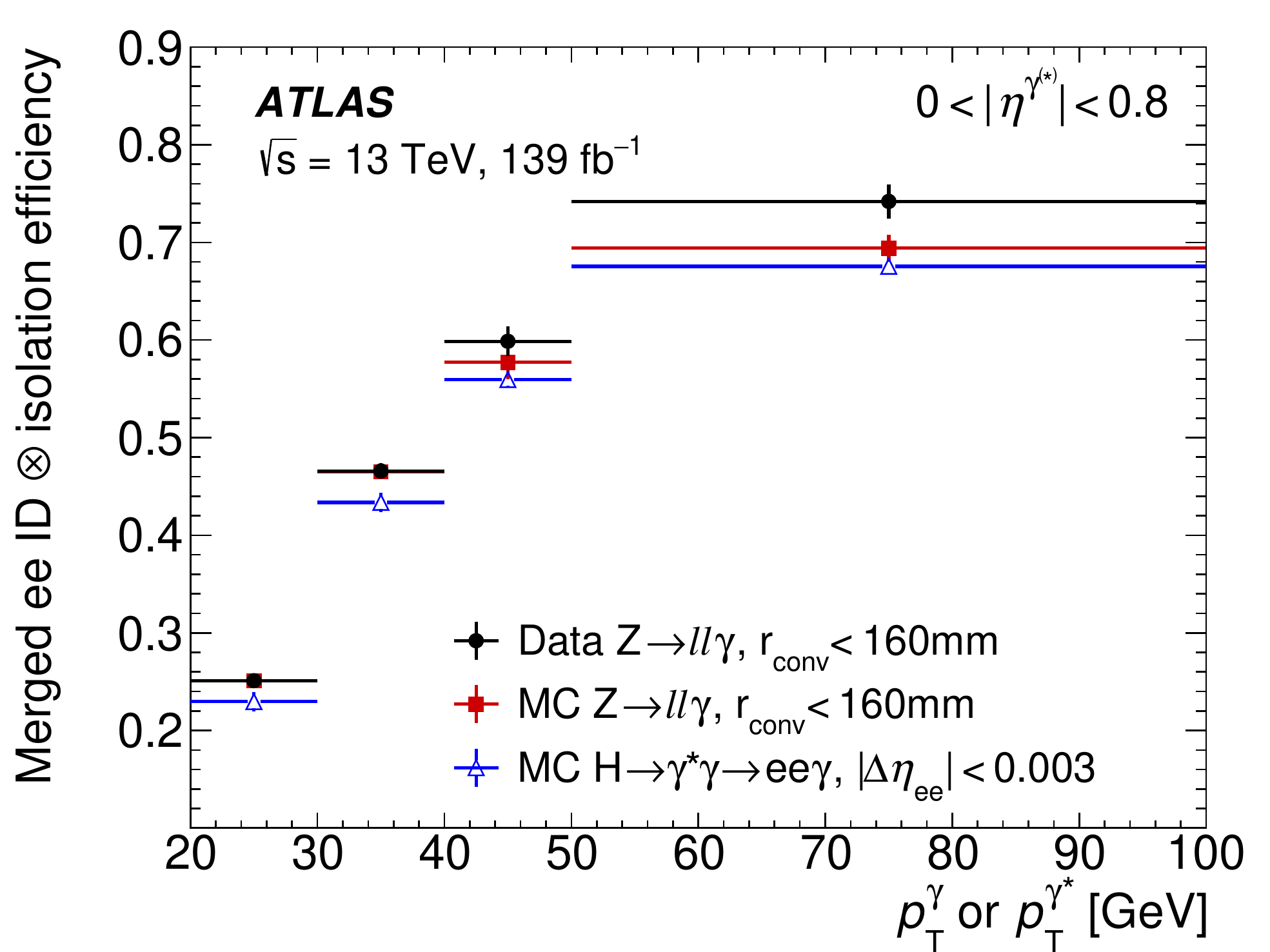} \\
(a) & (b) \\
\end{tabular}
\end{center}
\caption{(a) Ratio of reconstructed to true merged-$ee$ energy  in simulated $H \to \gamma^*\gamma \to ee \gamma$ events as a function of the true merged-$ee$ $\pT$ for several energy calibration techniques. The merged-$ee$ object is calibrated as a photon with a conversion radius of 30 mm (black circles, analysis choice), 100 mm (red squares), and 400 mm (blue upward triangles) or as an electron (purple downward triangles). (b) Combined merged-$ee$ identification and isolation efficiency extracted from $Z\to\ell\ell\gamma$ events with a photon that converts at a radius of $r_{\textrm{conv}}<160$~mm. The efficiencies are shown for photons with $|\eta|<0.8$ as a function of $\pT$.
Data (black circles) are compared with simulated $Z\to\ell\ell\gamma$ events (red squares). The resulting efficiencies are also compared with efficiencies in simulated $H \to \gamma^*\gamma \to ee \gamma$ events (blue triangles). On all points, the vertical error bars indicate the statistical uncertainties.
}
\label{fig:merged_perf}
\end{figure}

Jets are reconstructed from topological clusters~\cite{Aad:2016upy}
using the anti-$k_t$ algorithm~\cite{Cacciari:2008gp} with a radius parameter of 0.4 and are required to have $\pt > 20~\GeV$ and $|\eta|<4.4$.
Jets produced in pile-up interactions are suppressed by requiring that those with $\pt <
60~\GeV$ and $|\eta|<2.4$ pass a selection based on a jet vertex tagging algorithm~\cite{ATLAS-JVT}.

 
\section{Event selection}
\label{sec:evtselection}
Candidate \htoystary events must have at least one reconstructed photon, and at least one opposite-charge, same-flavour pair of leptons (muons or resolved electrons) or one merged-$ee$ object.
One of the muons (resolved electrons) in the lepton pair must have $\pT > 11~(13)~\GeV$, to match the \pt thresholds used in the dilepton triggers.
As discussed above, the merged-$ee$ \pt is required to be larger than 20~\GeV.
If the leading photon overlaps with one of the EM clusters (resolved electrons or merged-$ee$) forming the $\gamma^*$ candidate within $\Delta R <0.02$, the photon is discarded and the next-highest-\pT photon is considered.
No isolation requirement is applied to the subleading lepton (muon or resolved electron) as it is within the isolation cone of the leading lepton in the majority of events.
Additionally, if the subleading lepton (muon or electron) falls within the isolation cone of the leading lepton, it is not included in the calculation of the isolation variable.
 
Muon pairs are given the highest priority if there are lepton pairs of different
flavours in one event.
If no opposite-charge muon pair satisfying the requirements above is found, the electron-pair
candidates are considered.
The resolved electron pair or merged-$ee$ object with the highest vector-sum of the \pt of the associated ID tracks is selected.
 
In order to suppress events arising from FSR processes, events are rejected if the photon is within a $\Delta  R = 0.4$ cone around either of the selected leptons.
If the axis of a jet is within $\Delta R = 0.4$ of the photon or a muon, or within $\Delta R = 0.2$ of an electron (merged or resolved), the jet is discarded. However, if the electron--jet angular separation is $0.2< \Delta R <0.4$, the electron and therefore the event is discarded.

To suppress events involving  $Z$ boson decays, the dilepton invariant mass must satisfy $m_{\ell\ell}<30~\GeV$.
To remove $J/\psi\to\ell\ell$ production, events with a dimuon (dielectron) invariant mass in the range $2.9 <m_{\mu\mu}<3.3~\GeV$ ($2.5 <m_{ee}<3.5~\GeV$) are excluded.
Similarly, events with a dimuon (dielectron) invariant mass in the range $9.1<m_{\mu\mu}<10.6~\GeV$ ($8.0<m_{ee}<11.0~\GeV$) are rejected to avoid $\Upsilon(nS)\to\ell\ell$ contamination.
 
The invariant mass of the $\ell\ell\gamma$ system is required to satisfy $110 < \mllg < 160~\GeV$.
Additionally, both the photon and dilepton momenta must satisfy $\pT > 0.3\cdot \mllg$.

The selected events are classified into mutually exclusive categories, depending on the lepton types and event topologies.
The \vbfcat categories, which have the best expected signal-to-background ratio, but the lowest event count, are defined as follows. Events must contain at least two jets with $\pt>25~\GeV$.
If the leading or subleading jet is a forward jet (defined as a jet with $|\eta|>2.5$), it is required to have $\pt>30~\GeV$
to suppress jets originating from pile-up.
In addition, the invariant mass of the two leading jets, $m_{jj}$, must be greater than 500~\GeV, and
the pseudorapidity separation between the two leading jets, $|\Delta\eta_{jj}|$,
greater than 2.7.
The quantity $|\eta_{\llg}-0.5 (\eta_{j1}+\eta_{j2})|$~\cite{Rainwater:1996ud}
is required to be less than 2.0,
where $\eta_{\llg}$ is the pseudorapidity of the \llg system and $\eta_{j1}$ ($\eta_{j2}$)
is the pseudorapidity of the leading (subleading) jet.
The selected leptons and the photon must be separated from the two jets by $\Delta R>1.5$.
Additionally, the azimuthal separation between the $\llg$ system and the system formed by the two jets must be greater than 2.8. These jet variables are expected to have a different shape for the VBF signal and the background from QCD processes.

The $\ptt$ is defined as the component of the transverse momentum of the \llg system
that is perpendicular to the difference of the three-momenta of the dilepton system and the photon
candidate ($\ptt = |\vec{p}_\mathrm{T}^{\ell\ell\gamma} \times \hat{t} |$, where $\hat{t} = (\vec{p}_\mathrm{T}^{\ell\ell}-\vec{p}_\mathrm{T}^{\gamma})/|\vec{p}_\mathrm{T}^{\ell\ell}-\vec{p}_\mathrm{T}^{\gamma}|$).
This quantity is strongly correlated with the transverse momentum of the \llg system,
but has better experimental resolution~\cite{Ackerstaff:1997rc,VESTERINEN2009432}.
Events failing the \vbfcat selection, but having $\ptt>100~\GeV$, are assigned to the \highpttcat category, which has a lower expected signal-to-background ratio than the \vbfcat category, but is expected to have more events.
The \highpttcat category is expected to have an increased fraction of
\VBF and \VH events as these production modes lead on average to higher Higgs boson \pT than \ggF production.

The vast majority of selected events do not fall into the \vbfcat or \highpttcat categories discussed above, and are placed into the \lowpttcat category. The full list of all categories considered in the analysis, together with the expected yields for a 125.09~\GeV\ Higgs boson decaying into \llg, is shown in Table~\ref{tab:signal}.
The table also summarises the observed number of events in data in the $\mllg$ mass range of 110--160~\GeV.

\begin{table}
\centering
\caption{
Number of data events selected in each analysis category in the $\mllg$ mass range of 110--160~\GeV. In addition, the following numbers are given: number of \htoystary events in the smallest $\mllg$ window containing 90\% of the expected signal ($S_{90}$), the non-resonant background in the same interval ($B_{90}^N$) as estimated from fits to the data sidebands using the background models described in Section~\ref{sec:sigbkgmodel},
the resonant background in the same interval ($B_{\hyy}$),
the expected signal purity $f_{90} = S_{90}/(S_{90}+B_{90})$,
and the expected significance estimate defined as $Z_{90} = \sqrt{ 2( (S_{90}+B_{90})\,\ln(1+S_{90}/B_{90}) - S_{90}) }$ where $B_{90} = B_{90}^N+B_{\hyy}$. $B_{\hyy}$ is only relevant for the electron categories and is marked as ``--'' otherwise.
}
\vspace{0.2cm}
\begin{tabular}{l
r
@{\,\,\,\,\,\,\,\,\,}
S[table-format=2.1, table-number-alignment=center,round-mode=figures, round-precision=2]
@{\,\,\,\,\,\,\,\,\,}
S[table-format=4.1, table-number-alignment=center,round-mode=figures, round-precision=2]
@{\,\,\,\,\,\,\,\,\,}
S[table-format=2.1, table-number-alignment=center,round-mode=figures, round-precision=1]
@{\,\,\,\,\,\,\,\,\,}
S[table-format=2.1, table-number-alignment=center,round-mode=figures, round-precision=2]
@{\,\,\,\,\,\,\,\,\,}
S[table-format=1.2, table-number-alignment=center,round-mode=figures, round-precision=2]
}
\toprule
Category          & {Events} & {$S_{90}$} & {$B_{90}^N$} & {$B_{\hyy}$} & {$f_{90}$~[\%]} & {$Z_{90}$} \\ \midrule
\vbfee            &     10   & {\pho\numRF{  0.41}{1}} &                 1.62    &   0.009     &             20.0           & {\pho\numRF{  0.3}{1}} \\
\vbfeemerged      &     15   & {\pho\numRF{  0.75}{1}} &                 1.98      &   0.065   &             26.9           & {\pho\numRF{ 0.5}{1}} \\
\vbfmumu          &     33   &               1.25      &                 5.85      &  \phantom{0.8}\texttt{--}   &             17.7           & {\pho\numRF{ 0.5}{1}} \\
\highpttee        &     86   &               1.11      &                11.81     &   0.024    & {\pho\numRF{ 8.6}{1}\phdo} & {\pho\numRF{ 0.3}{1}} \\
\highptteemerged  &    162   &               2.48      &                17.59      &  0.18    &             12.2           & {\pho\numRF{ 0.6}{1}} \\
\highpttmumu      &    210   &               4.01      &                33.82     &  \phantom{0.8}\texttt{--}   &             10.6          & {\pho\numRF{ 0.7 }{1}}\\
\lowpttee         &   3713   &              21.64      & {\pho\numRF{  728.9 }{3}\phdo} &   0.489     &      2.9           & {\pho\numRF{ 0.8 }{1}}\\
\lowptteemerged   &   5103   &              29.04      & {\pho\numRF{  941.55 }{3}\phdo} &  1.887   &         3.0           & {\pho\numRF{ 1.0 }{2}}\\
\lowpttmumu       &   9813   &              60.71      &     {\numRF{ 1745.73 }{3}\phdo} &  \phantom{0.8}\texttt{--}  &         3.4           & {\pho\numRF{ 1.4}{2}} \\
\bottomrule
\end{tabular}
\label{tab:signal}
\end{table}
 
\section{Signal and background modelling}
\label{sec:sigbkgmodel}
To extract the observed signal yield, parametric functions are chosen to model the \llg invariant mass distributions of signal and background in each analysis category. A combined model is built from these functions and fit to the selected data in the \mllg range of 110$-$160~\gev, simultaneously in all categories.
 
The signal model, including its parameters, is obtained by fitting a double-sided Crystal Ball function (DSCB)~\cite{dscb1,dscb2} to the \mllg distribution obtained from the \hgsg samples described in Section~\ref{sec:datamcsamples} after applying the event selection and categorisation from Section~\ref{sec:evtselection}.
The DSCB function features a Gaussian core and asymmetric power-law tails.
A shift of +0.09~\gev\ is applied to the mean of the Gaussian function to account for the fact that the sample assumes a Higgs boson mass of 125~\gev.
The effective signal mass resolution, defined as half the width containing 68\% of the signal events, depends on the category and lies between 1.6~\gev\ (\highptteemergedcat category) and 2.2~\gev\ (\lowptteecat category).

The \hgg process, which contributes as background to the electron categories through converted photons, is modelled using the same functions and parameters as the signal in the respective categories, and is normalised to the predicted SM yield. The parameterisations are compatible with the statistically limited distributions obtained from simulated \hgg events.
This background contribution is relatively small ($<$2.5\% and $<$7\% of the expected \hgsg signal in the $ee$-resolved channel and $ee$-merged channel, respectively) and is taken into account in the fitting procedure.
 
The non-resonant background is also estimated with parametric functions. The background normalisation and function parameters are allowed to float in the fit to data. The chosen functional forms are based on background templates that are built taking into account the contributions of different processes to the background.
The function choice is performed separately for each category.
 
The dominant part of the background originates from the non-resonant \llg process. There is also a smaller background from events with misidentified photons, electrons, or muons. To estimate the fraction of events with a misidentified photon, a control region is formed using the signal selection, but dropping the photon isolation requirement and using it as a discriminating variable in a template fit. A background template enriched in events with misidentified photons is built by inverting the photon identification selection, with the prompt-photon contamination removed by subtracting its distribution found in simulated events.
The template normalisation is obtained in the background-dominated sideband of the photon isolation distribution for each category separately. In the signal region, about 10\% of all selected events have a misidentified photon, independently of the category.
The fraction of events with a misidentified electron or muon that is in fact a hadronic jet, a lepton from heavy-flavour decays or an electron from a photon conversion, is estimated in a similar manner, using a control region in which the isolation selection is dropped for the softer lepton. The estimated fractions of events with misidentified leptons are 4\% in the \lowpttmumucat category, 2\% in the \lowptteemergedcat category, and 30\% in the \lowptteecat category.
Events in the last category are separated into two groups depending on the angular distance between the electrons, as two populations with different misidentified-electron rates and mass distributions were found.
The numbers in the other categories are extracted as well, but suffer from fairly large statistical uncertainties.

The invariant mass template for the non-resonant \llg background is built from the simulated events described in Section~\ref{sec:datamcsamples}. The invariant mass templates for events with misidentified objects are obtained from background-dominated control regions, and scaled by the yields derived above.
Reasonable agreement is observed between the templates containing the sum of all backgrounds and the sidebands of the \mllg distributions in data (105$-$120~\gev\ and 130$-$160~\gev).

The choice of fit function for the non-resonant background is made in each category using signal-plus-background fits to the constructed background-only template by measuring the bias associated with each function, expressed as the number of fitted signal events, and choosing the function with the smallest number of degrees of freedom that satisfies the bias criteria described below.
 
The functional forms used to model the background are selected from the following:
exponential ($\mathrm{e}^{\alpha \mllg }$), exponential of a second-order polynomial ($\mathrm{e}^{\alpha\mllg  + \beta\mllg^2 }$), and a  power-law function ($\mllg^{\alpha}$), where $\alpha$ and $\beta$ are free parameters.
Signal hypotheses with \mh ranging from 121~\GeV\ to 129~\GeV\ are tested in steps of 1~\gev, and the fit bias is evaluated as the absolute maximum bias over this range (referred to as the `spurious signal' in the following), similar to what is done in Ref.~\cite{HIGG-2012-27}.
A function passes the test if the spurious signal is less than 10\% of the expected number of \htoystary events or less than 20\% of the statistical uncertainty of the fitted spurious signal due to the expected number of background events.
To account for statistical fluctuations in the background template, the criteria are relaxed by the statistical uncertainty due to the template~\cite{HIGG-2016-21}.
Furthermore, in a background-only fit of the template in the mass range $110 < \mllg < 160~\GeV$, the function must pass a $\chi^2$ test with a probability larger than 1\%. As described in Ref.~\cite{HIGG-2016-21}, an additional check is performed by fitting both the chosen function and a function belonging to the same family with one more degree of freedom to the sidebands of the \mllg distribution in data, to ensure the data do not prefer the more complex function.
Following the procedure described above, the power-law function is selected for all categories, except for
the \lowpttmumucat and \lowptteemergedcat categories, which are best described by the second-order exponential polynomial, and the \vbfeecat category, for which the exponential function is chosen.
 
An unbinned extended likelihood function is formed from the product of each category's parameterised signal-plus-background probability density function.
Systematic uncertainties are considered in the form of nuisance parameters with Gaussian or log-normal constraints. They are correlated across all categories, except for the spurious-signal uncertainties.
The latter are implemented for each category as a signal-like component with a yield parameter that is constrained by a Gaussian function, centred at zero, with a width corresponding to the estimated spurious signal.
The Higgs boson mass is set to 125.09~$\pm$~0.24~\GeV~\cite{HIGG-2014-14}.
 
The parameter of interest is the signal strength $\mu$, which is defined as the ratio of the measured signal yield to the SM expectation, taking into account the uncertainties on the latter.
The corresponding profile likelihood ratio is maximised to extract the best-fit $\mu$~\cite{stat}.
A possible excess over the background-only hypothesis is quantified by a $p$-value using the profile likelihood ratio, evaluated for a vanishing $H \rightarrow\ell\ell\gamma$  branching ratio, as a test statistic.
For all results, the asymptotic approximation is used~\cite{stat}.
Cross-checks of the best-fit $\mu$ and the $p$-value extracted with pseudo-experiments show good agreement with those obtained using the asymptotic approximation.

The measurement of the signal strength can be converted into a measurement of
the Higgs boson production cross-section times the $H \rightarrow\ell\ell\gamma$ branching ratio in the fiducial region \mll $<$ 30~\GeV.
In this case, the included theory uncertainties are adjusted as discussed in Section~\ref{sec:sys}, and the acceptance is extracted with the simulated samples described in Section~\ref{sec:datamcsamples}.

 
\section{Systematic uncertainties}
\label{sec:sys}
The total observed systematic uncertainty in the signal strength measurement is 11\%, which is about 35\% of the size of the statistical uncertainty. Therefore, systematic uncertainties do not play a dominant role in this analysis.
 
The dominant experimental systematic uncertainties are due to the estimated biases in the fitted signal events (spurious signal, see Section~\ref{sec:sigbkgmodel}). The corresponding uncertainty in the observed signal strength amounts to 6.1\%.
Other non-negligible systematic uncertainties relate to photon and lepton identification efficiencies in the simulated signal samples, in particular for merged-$ee$ objects, as well as the energy/momentum calibration (see Section~\ref{sec:objselection}). Including jet uncertainties, which have a much smaller impact, these add up to 4.0\%.
 
The uncertainty in the combined 2015--2018 integrated luminosity is 1.7\%~\cite{ATLAS-CONF-2019-021}, obtained using the LUCID-2 detector~\cite{LUCID2} for the primary luminosity measurements.
Uncertainties in the pile-up modelling contribute 1.7\%.
Uncertainties in the estimate of the \hgg background have a small impact of 0.7\%. They include uncertainties in the photon--electron fake rate, and the uncertainties in the $\sigma (H) \times \mathcal{B}(H \rightarrow \gamma \gamma)$ measurement~\cite{HIGG-2016-21}.
 
The assumed uncertainty in the \hysy branching ratio contributes 5.8\%. The choice of QCD scales impacts the number and distribution of signal events in the different categories and is evaluated for the \ggF, \VBF, and \VH production modes using a scheme similar to the one discussed in Ref.~\cite{HIGG-2016-21}. The corresponding uncertainty in the measured signal strength is 4.7\%. Uncertainties in the PDF are evaluated using the eigenvectors of the PDF4LHC15 PDF set~\cite{Butterworth:2015oua} and have a smaller effect of 2.3\%. A conservative uncertainty of 50\% is assigned to the normalisation of the \ttH, \ggZH, and \bbH production modes, as no dedicated \hysy samples are available, with an impact of 0.8\%. Parton shower uncertainties contribute only 0.3\%.
 
For the $\sigma (H) \times \mathcal{B}(H \rightarrow\ell\ell\gamma$) measurement, the effect of the theory uncertainties is reduced to 1.1\% for the QCD scale and 0.9\% for the PDF uncertainty, as only acceptance effects are considered, whereas the uncertainties in the predicted cross-sections and branching ratio are not applicable to this measurement.


\section{Results}
\label{sec:result}
The \mllg distributions of the selected events and the result of the global fit of the parametric signal-plus-background models to the data are shown in Figure~\ref{fig:fit_per_category} for each event category.

\begin{figure}[!htb]
\begin{center}
\begin{tabular}{ccc}
\includegraphics[width=0.33\textwidth]{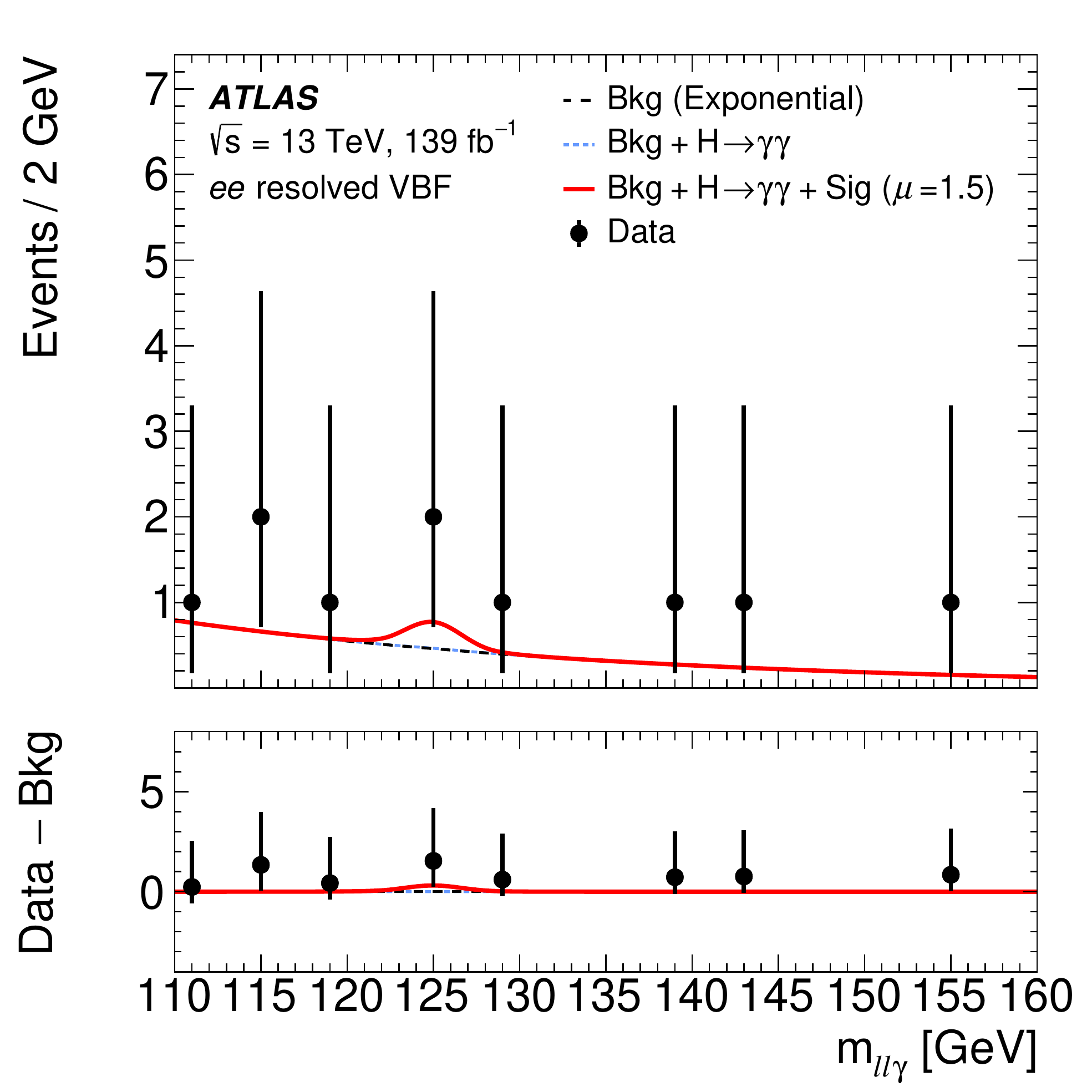} &  \hspace{-0.4cm}
\includegraphics[width=0.33\textwidth]{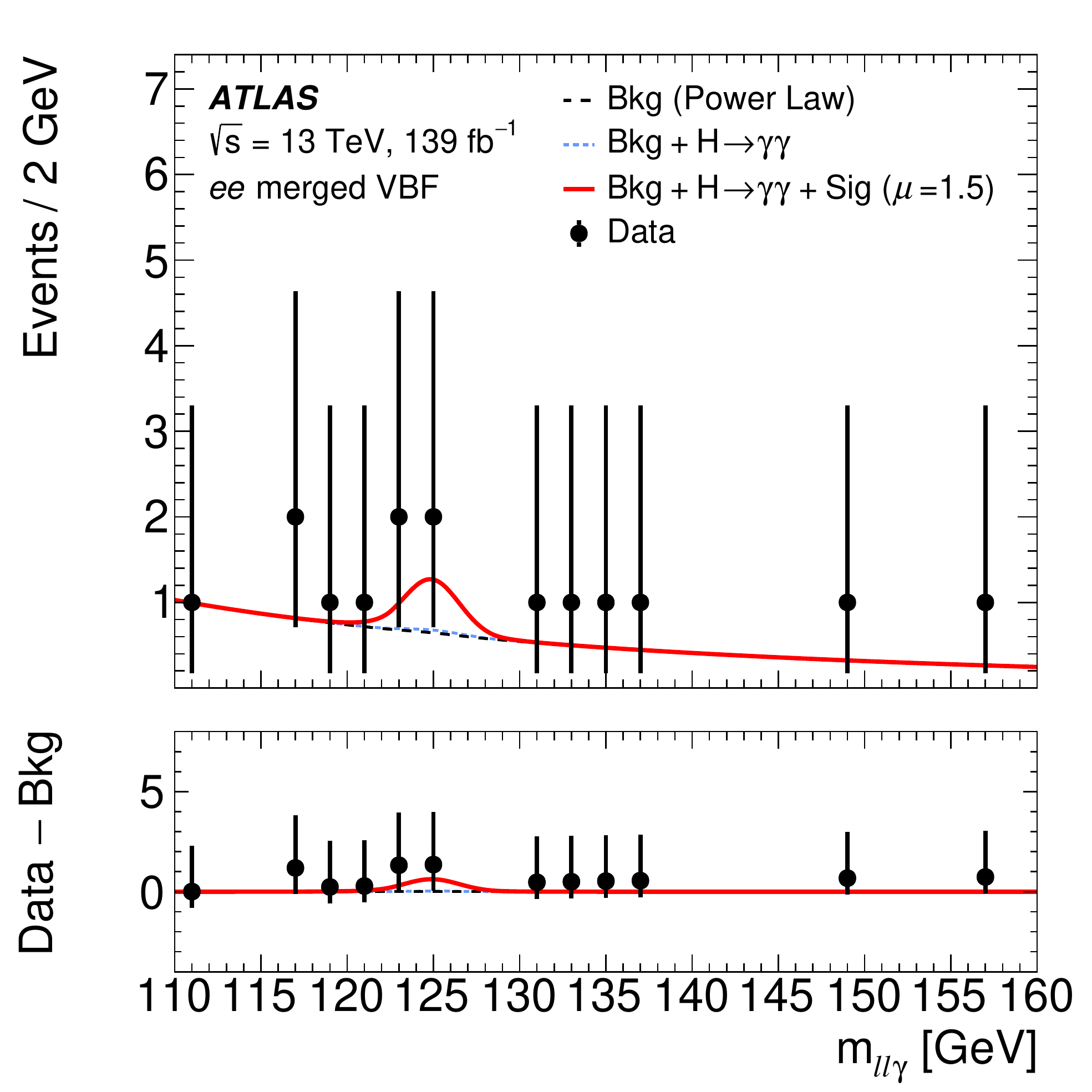} & \hspace{-0.4cm}
\includegraphics[width=0.33\textwidth]{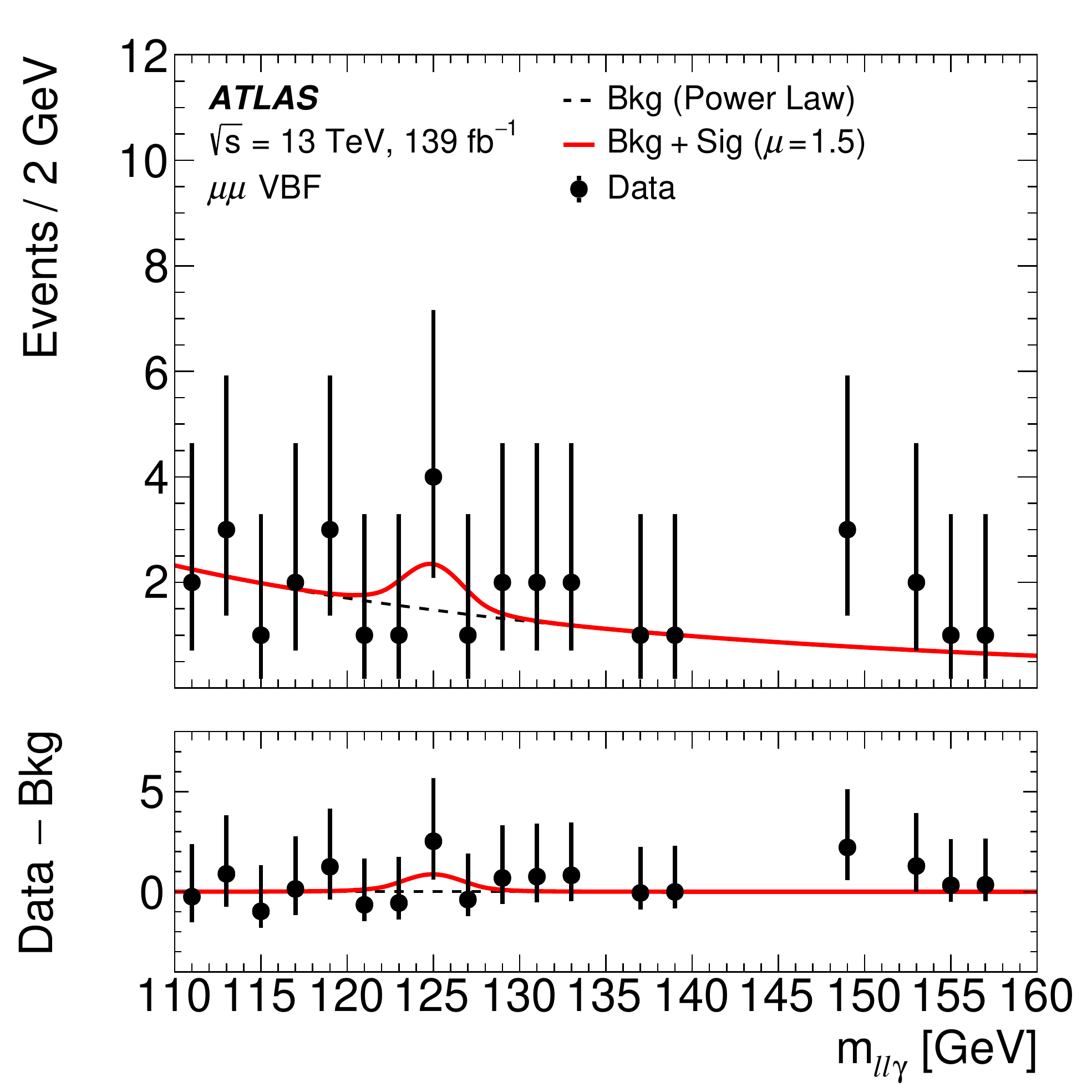} \\
(a) & (b) & (c)\\
\includegraphics[width=0.33\textwidth]{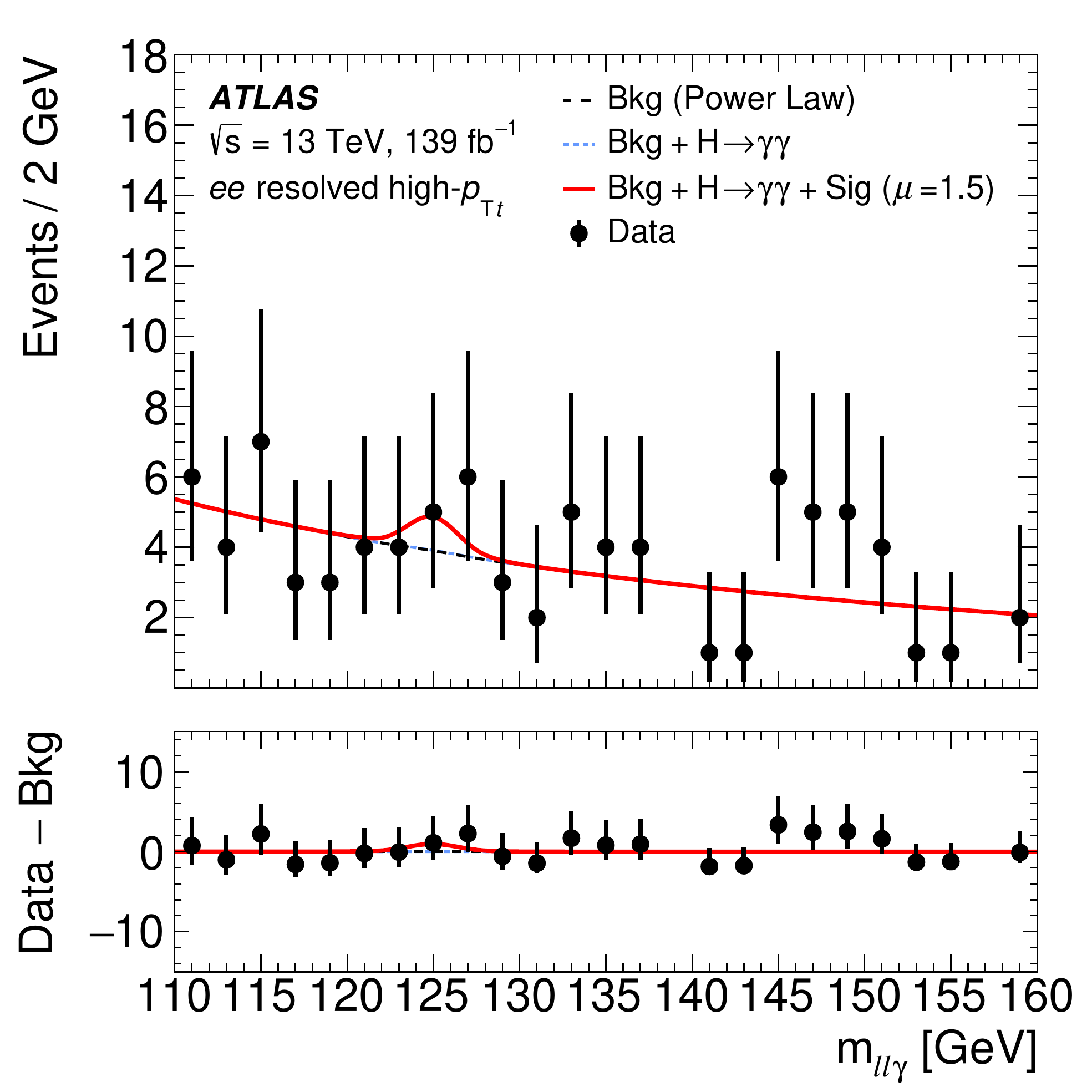} &  \hspace{-0.4cm}
\includegraphics[width=0.33\textwidth]{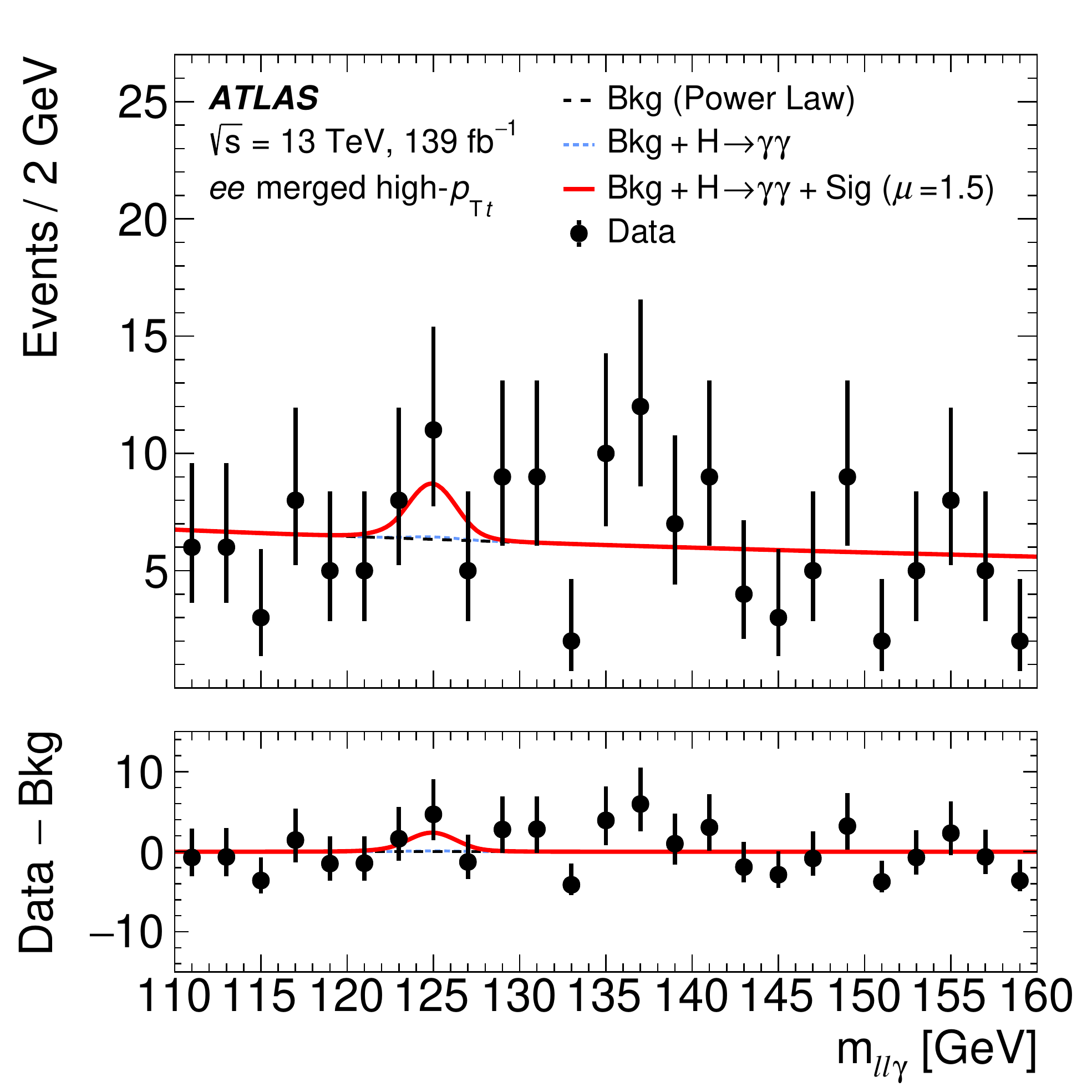} & \hspace{-0.4cm}
\includegraphics[width=0.33\textwidth]{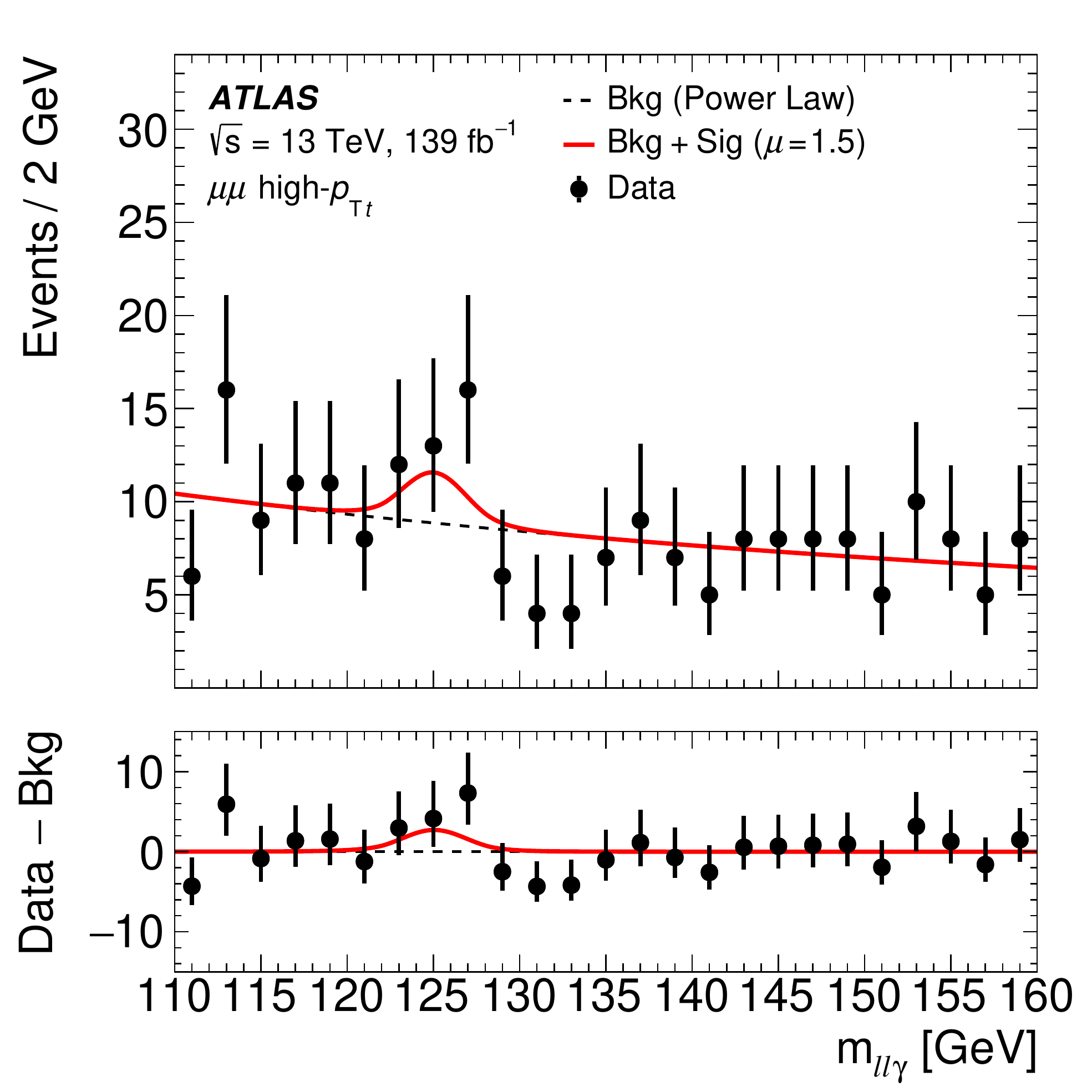} \\
(d) & (e) & (f)\\
\includegraphics[width=0.33\textwidth]{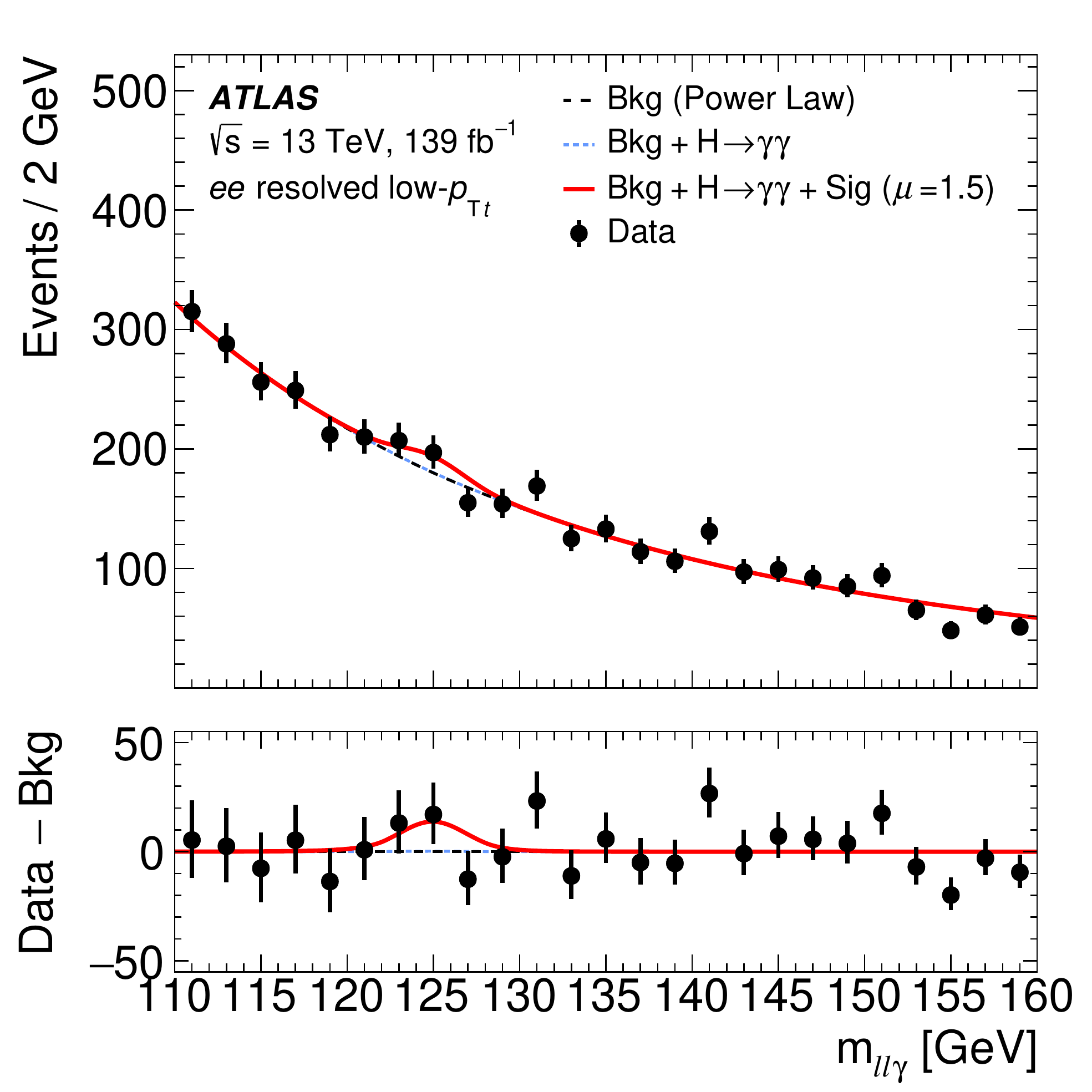} &  \hspace{-0.4cm}
\includegraphics[width=0.33\textwidth]{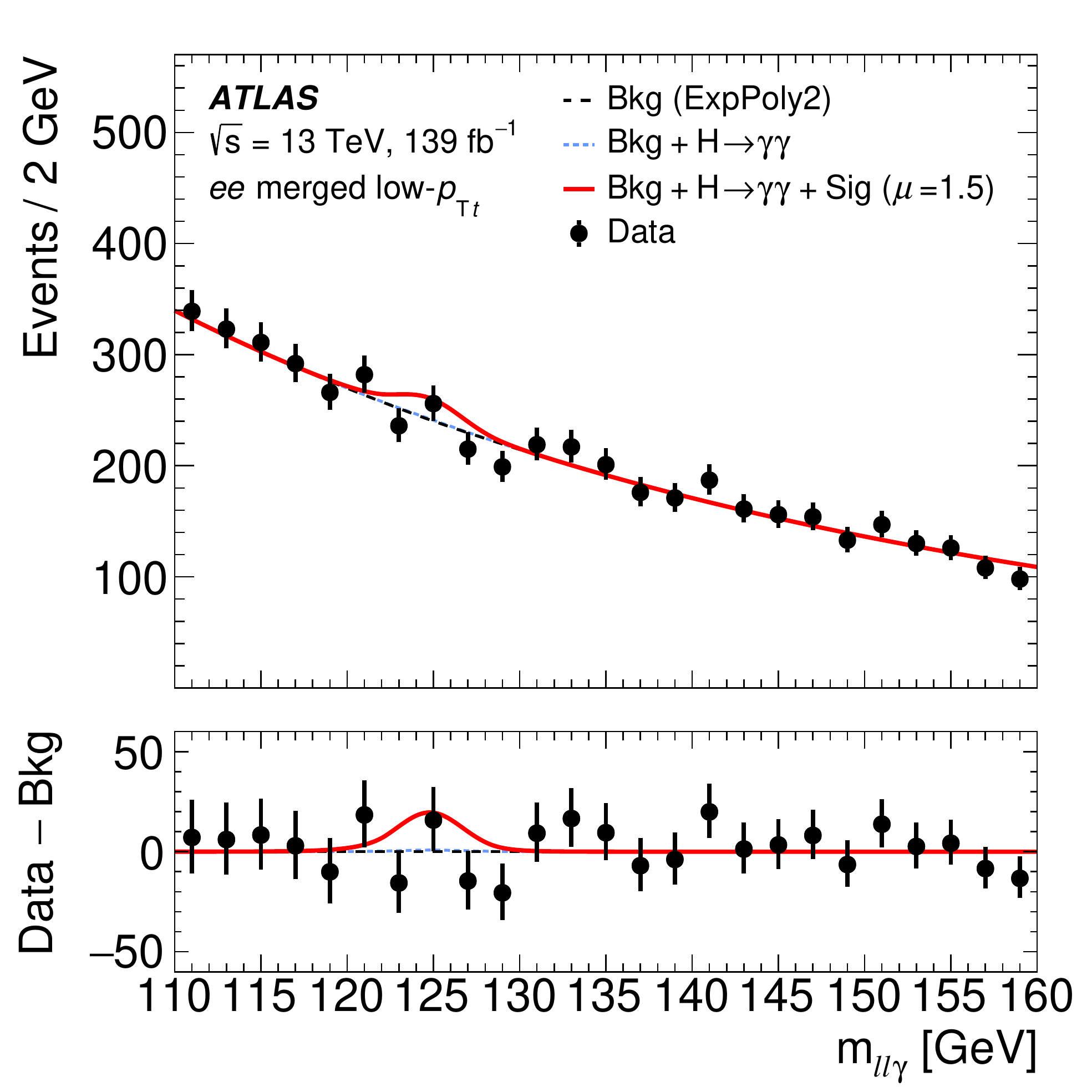} &  \hspace{-0.4cm}
\includegraphics[width=0.33\textwidth]{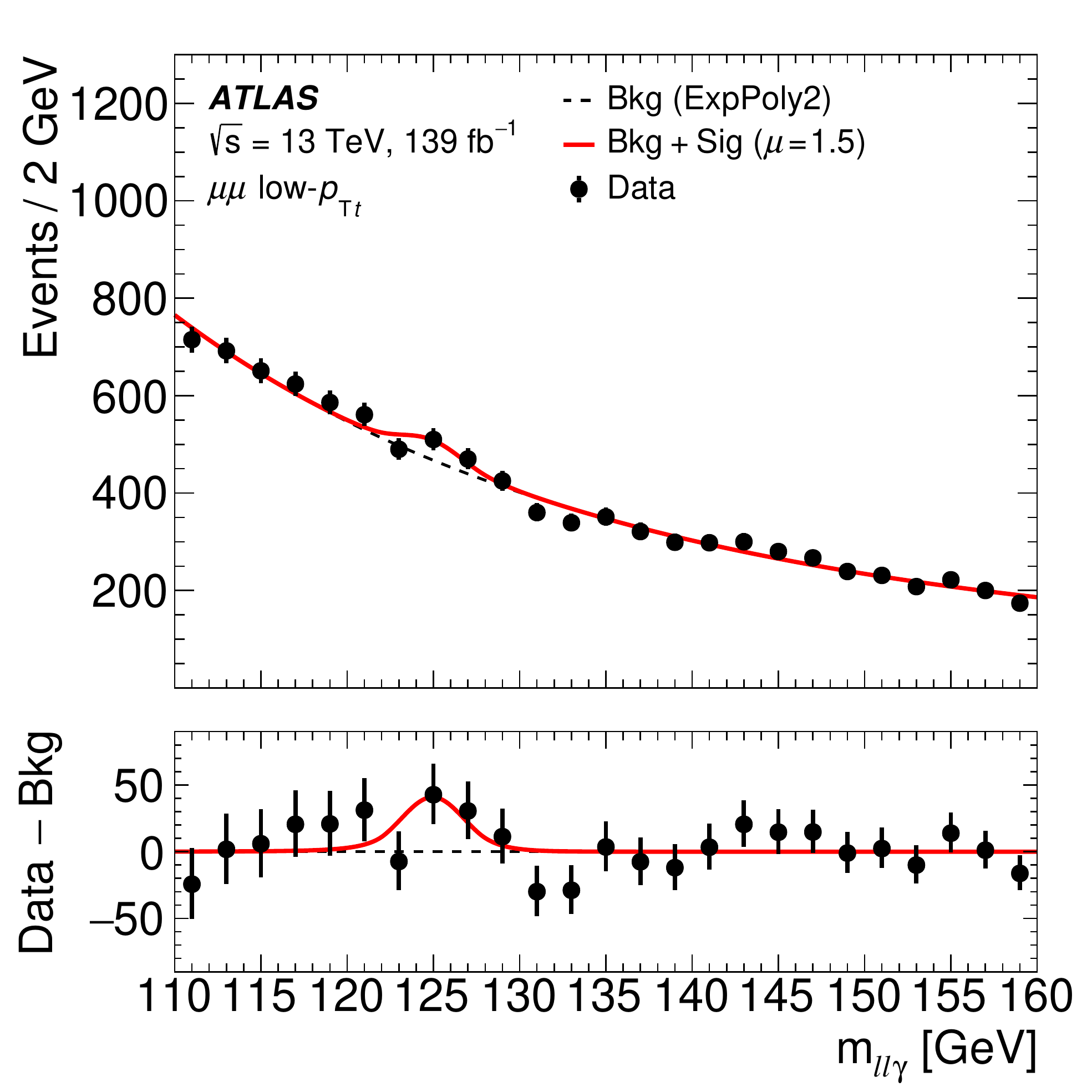} \\
(g) & (h) & (i)\\
\end{tabular}
\end{center}
\caption{\mllg distributions of the selected events and the results of the global fit, for the \vbfcat categories (a, b, c), the \highpttcat categories (d, e, f), and the \lowpttcat categories (g, h, i). The $ee$-resolved categories are shown in the left column, the $ee$-merged categories in the middle and the $\mu\mu$ categories in the right column. The data are shown as the black circles with statistical uncertainties.
The red curve shows the combined signal-plus-background model, the dashed black line shows the model of the non-resonant background component and the dotted blue line denotes the sum of the non-resonant background and the resonant \hgg background. The curves are obtained from the fit, i.e.\ they include the best-fit values of the parameter of interest and the nuisance parameters, including the spurious signal.
The bottom panel shows the residuals of the data with respect to the non-resonant background component of the signal-plus-background fit.
}
\label{fig:fit_per_category}
\end{figure}

The best-fit value of the signal-strength parameter is $\mu=\muobs$, while the corresponding expected SM value is $\mu_\mathrm{exp}=\muexp$. The best-fit signal strength in the muon (electron) channel, obtained from a  fit with two separate signal-strength parameters, is $\mu_{\mu\mu}=\muobsmumu$ ($\mu_{ee}=\muobsee$). Figure~\ref{fig:fit_individual} shows the results of the fit when the signal strength in each category is allowed to float independently. As anticipated in Table~\ref{tab:signal}, the \lowpttcat categories, especially in the $\mu\mu$ channel, have the smallest uncertainties.
It can be seen that all categories yield results that are consistent with each other and with the result of the single-$\mu$ fit.
 
For illustration, Figure~\ref{fig:fit_weighted} shows the \mllg distribution, with every data event reweighted by a category-dependent weight $\ln{(1+S_{90}/B_{90})}$, where $S_{90}$ is the number of signal events in the smallest window containing 90\% of the expected signal and $B_{90}$ is the expected number of background events in the same window, which consists of the resonant \hyy\ background as well as the non-resonant background. The latter is estimated from fits to the data sidebands using the background models.

The observed (expected) significance over the background-only hypothesis for a Higgs boson with a mass of 125.09~\GeV\ is \sigobs (\sigexp).
 
The Higgs boson production cross-section times the $H \rightarrow\ell\ell\gamma$ branching ratio for \mll $<$ 30~\GeV\ is determined to be \totXSobs.

\begin{figure}[htb]
\begin{center}
\includegraphics[width=0.45\textwidth]{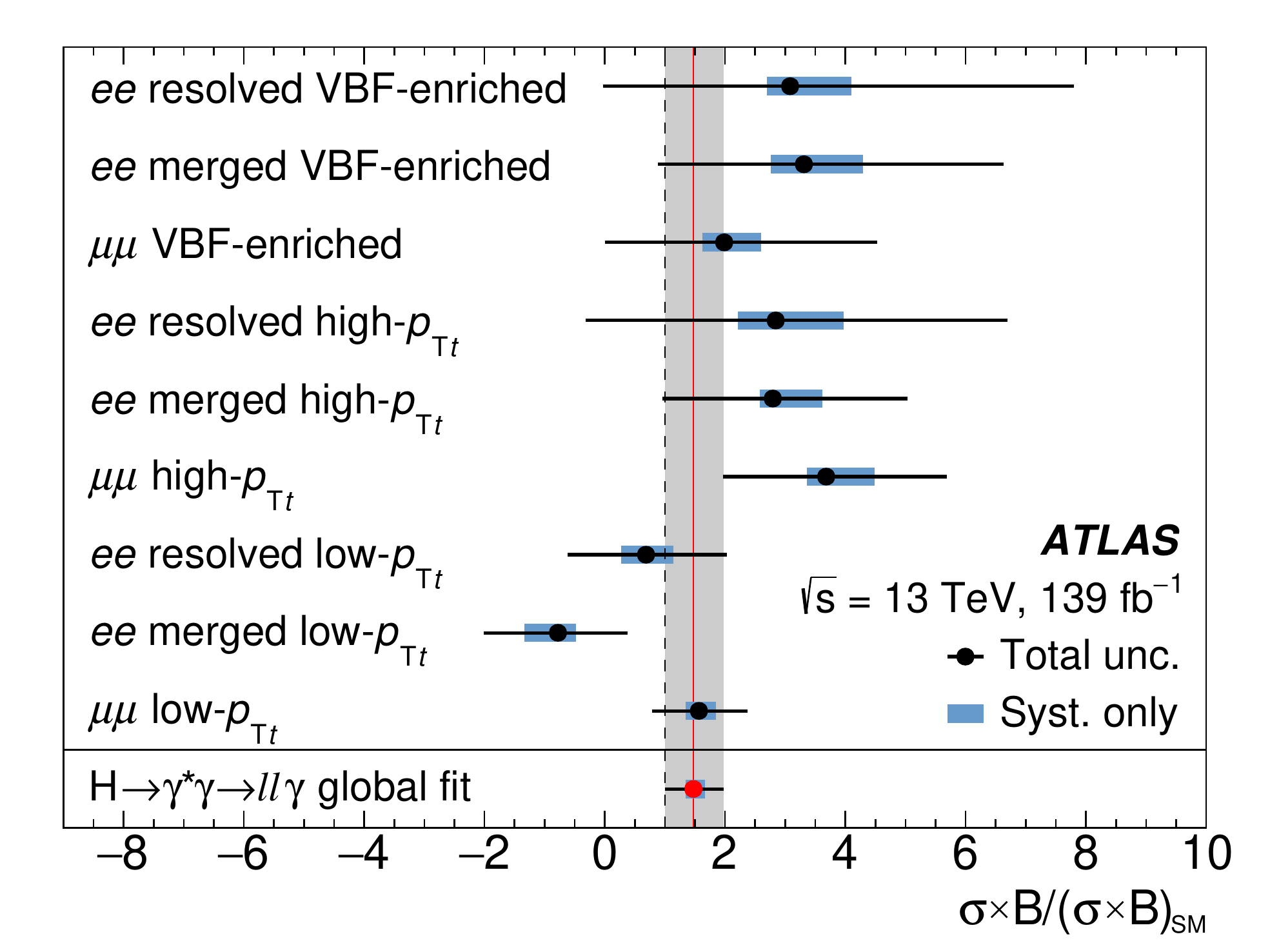}
\end{center}
\caption{Best-fit values of the signal-strength parameters for all event categories, in a fit where the signal strength in each category is allowed to float independently (black circles), compared with the result of the global fit (red circle and line) including its total uncertainty (grey band).
}
\label{fig:fit_individual}
\end{figure}

\begin{figure}[htb]
\begin{center}
\includegraphics[width=0.45\textwidth]{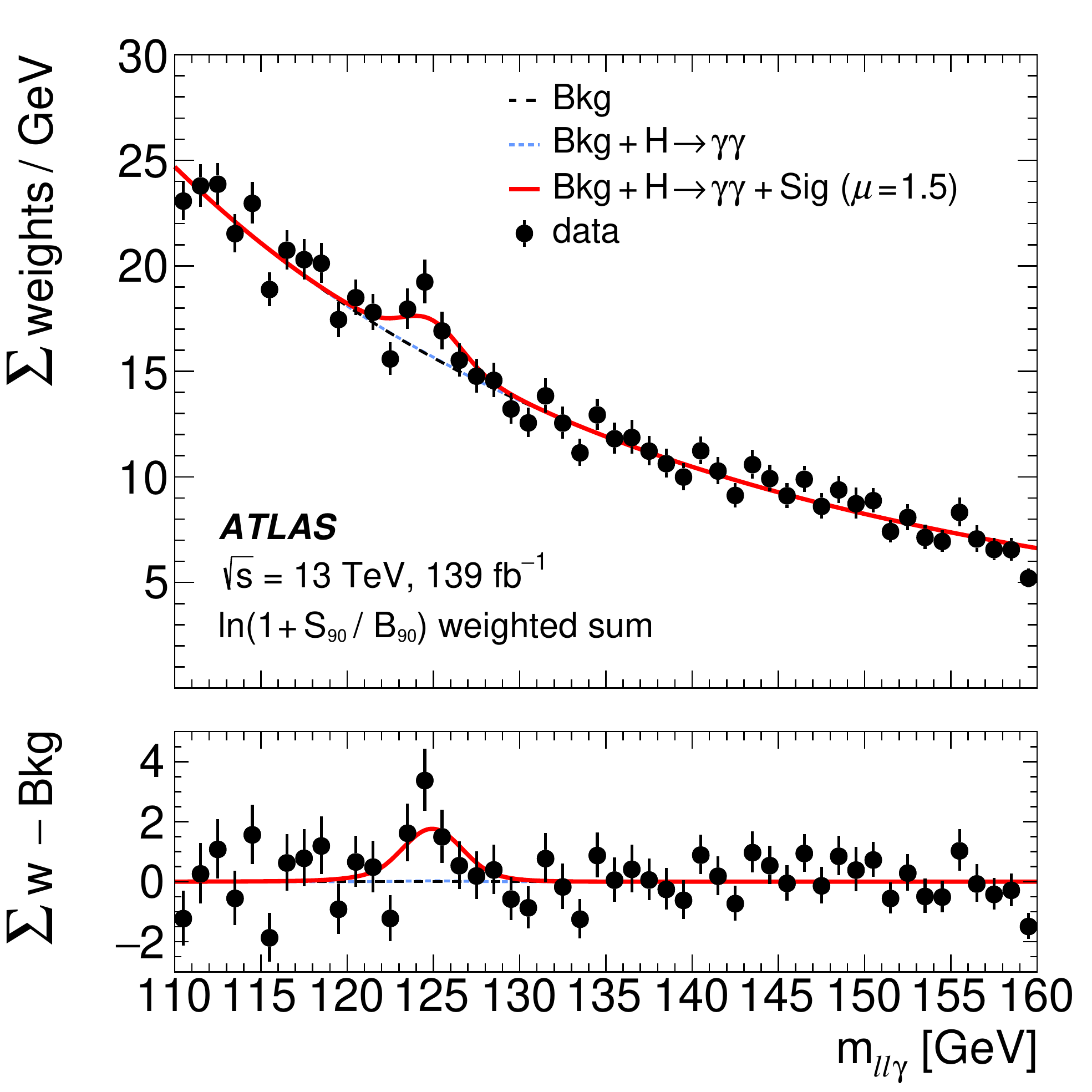}
\end{center}
\caption{\mllg distribution, with every data event reweighted by a category-dependent weight, $\ln{(1+S_{90}/B_{90})}$, where $S_{90}$ is the number of signal events in the smallest window containing 90\% of the expected signal, and $B_{90}$ is the expected number of background events in the same window, estimated from fits to the data sidebands using the background models. The data are shown as the black circles with statistical uncertainties. The parameterised signal and backgrounds are also added up with the category-dependent weight.
The red curve shows the combined signal-plus-background model when fitting all analysis categories simultaneously, the dashed black line shows the model of the non-resonant background component and the dotted blue  line denotes  the sum of the non-resonant background and the resonant \hgg background. The curves are obtained from the fit, i.e.\ they include the best-fit values of the parameter of interest and the nuisance parameters, including the spurious signal.
The bottom panel shows the residuals of the data with respect to the non-resonant background component of the signal-plus-background  fit.
}
\label{fig:fit_weighted}
\end{figure}

\FloatBarrier

\section{Conclusion}
\label{sec:conclusion}
A search for the Higgs boson decaying into a low-mass pair of electrons or muons and a photon is presented.
The analysis is performed using a data set recorded by the ATLAS experiment at the LHC with proton--proton collisions at a centre-of-mass energy of 13~\TeV,
corresponding to an integrated luminosity of 139~fb$^{-1}$.
For a Higgs boson with a mass of 125.09~\GeV\ and \mll $<$ 30~\GeV, evidence for the \hllg process is found with a significance of \sigobs over the background-only hypothesis, compared to an expected significance of \sigexp.
The best-fit value of the signal-strength parameter, defined as the ratio of the observed signal yield to the signal yield expected in the Standard Model, is $\mu = $ \muobsShort. The Higgs boson production cross-section times the $H \rightarrow\ell\ell\gamma$ branching ratio for \mll $<$ 30~\GeV\ is determined to be \totXSobsShort.
This result constitutes the first evidence for the decay of the Higgs boson into a pair of leptons and a photon, an important step towards probing Higgs boson couplings in this rare decay channel.

\section*{Acknowledgements}

We thank CERN for the very successful operation of the LHC, as well as the
support staff from our institutions without whom ATLAS could not be
operated efficiently.
 
We acknowledge the support of ANPCyT, Argentina; YerPhI, Armenia; ARC, Australia; BMWFW and FWF, Austria; ANAS, Azerbaijan; SSTC, Belarus; CNPq and FAPESP, Brazil; NSERC, NRC and CFI, Canada; CERN; ANID, Chile; CAS, MOST and NSFC, China; COLCIENCIAS, Colombia; MSMT CR, MPO CR and VSC CR, Czech Republic; DNRF and DNSRC, Denmark; IN2P3-CNRS and CEA-DRF/IRFU, France; SRNSFG, Georgia; BMBF, HGF and MPG, Germany; GSRT, Greece; RGC and Hong Kong SAR, China; ISF and Benoziyo Center, Israel; INFN, Italy; MEXT and JSPS, Japan; CNRST, Morocco; NWO, Netherlands; RCN, Norway; MNiSW and NCN, Poland; FCT, Portugal; MNE/IFA, Romania; JINR; MES of Russia and NRC KI, Russian Federation; MESTD, Serbia; MSSR, Slovakia; ARRS and MIZ\v{S}, Slovenia; DST/NRF, South Africa; MICINN, Spain; SRC and Wallenberg Foundation, Sweden; SERI, SNSF and Cantons of Bern and Geneva, Switzerland; MOST, Taiwan; TAEK, Turkey; STFC, United Kingdom; DOE and NSF, United States of America. In addition, individual groups and members have received support from BCKDF, CANARIE, Compute Canada, CRC and IVADO, Canada; Beijing Municipal Science \& Technology Commission, China; COST, ERC, ERDF, Horizon 2020 and Marie Sk{\l}odowska-Curie Actions, European Union; Investissements d'Avenir Labex, Investissements d'Avenir Idex and ANR, France; DFG and AvH Foundation, Germany; Herakleitos, Thales and Aristeia programmes co-financed by EU-ESF and the Greek NSRF, Greece; BSF-NSF and GIF, Israel; La Caixa Banking Foundation, CERCA Programme Generalitat de Catalunya and PROMETEO and GenT Programmes Generalitat Valenciana, Spain; G\"{o}ran Gustafssons Stiftelse, Sweden; The Royal Society and Leverhulme Trust, United Kingdom.
 
The crucial computing support from all WLCG partners is acknowledged gratefully, in particular from CERN, the ATLAS Tier-1 facilities at TRIUMF (Canada), NDGF (Denmark, Norway, Sweden), CC-IN2P3 (France), KIT/GridKA (Germany), INFN-CNAF (Italy), NL-T1 (Netherlands), PIC (Spain), ASGC (Taiwan), RAL (UK) and BNL (USA), the Tier-2 facilities worldwide and large non-WLCG resource providers. Major contributors of computing resources are listed in Ref.~\cite{ATL-SOFT-PUB-2020-001}.
 

\printbibliography

\clearpage 
 
\begin{flushleft}
\hypersetup{urlcolor=black}
{\Large The ATLAS Collaboration}

\bigskip

\AtlasOrcid[0000-0002-6665-4934]{G.~Aad}$^\textrm{\scriptsize 101}$,    
\AtlasOrcid[0000-0002-5888-2734]{B.~Abbott}$^\textrm{\scriptsize 127}$,    
\AtlasOrcid[0000-0002-7248-3203]{D.C.~Abbott}$^\textrm{\scriptsize 102}$,    
\AtlasOrcid[0000-0002-2788-3822]{A.~Abed~Abud}$^\textrm{\scriptsize 36}$,    
\AtlasOrcid[0000-0002-1002-1652]{K.~Abeling}$^\textrm{\scriptsize 53}$,    
\AtlasOrcid[0000-0002-2987-4006]{D.K.~Abhayasinghe}$^\textrm{\scriptsize 93}$,    
\AtlasOrcid[0000-0002-8496-9294]{S.H.~Abidi}$^\textrm{\scriptsize 29}$,    
\AtlasOrcid[0000-0002-8279-9324]{O.S.~AbouZeid}$^\textrm{\scriptsize 40}$,    
\AtlasOrcid{N.L.~Abraham}$^\textrm{\scriptsize 155}$,    
\AtlasOrcid[0000-0001-5329-6640]{H.~Abramowicz}$^\textrm{\scriptsize 160}$
% ...
\bigskip
\\

% ...
$^{29}$Physics Department, Brookhaven National Laboratory, Upton NY; United States of America.\\
% ...
$^{36}$CERN, Geneva; Switzerland.\\
% ...
$^{40}$Niels Bohr Institute, University of Copenhagen, Copenhagen; Denmark.\\
% ...
$^{53}$II. Physikalisches Institut, Georg-August-Universit\"{a}t G\"ottingen, G\"ottingen; Germany.\\
% ...
$^{93}$Department of Physics, Royal Holloway University of London, Egham; United Kingdom.\\
% ...
$^{101}$CPPM, Aix-Marseille Universit\'e, CNRS/IN2P3, Marseille; France.\\
$^{102}$Department of Physics, University of Massachusetts, Amherst MA; United States of America.\\
% ...
$^{127}$Homer L. Dodge Department of Physics and Astronomy, University of Oklahoma, Norman OK; United States of America.\\
% ...
$^{155}$Department of Physics and Astronomy, University of Sussex, Brighton; United Kingdom.\\
% ...
$^{160}$Raymond and Beverly Sackler School of Physics and Astronomy, Tel Aviv University, Tel Aviv; Israel.\\
% ...

\end{flushleft}


\end{document}